\documentclass[12pt,english,longbibliography,nofootinbib,superscriptaddress,12pt,sort&compress,showkeys]{revtex4-1}
\usepackage{ae,aecompl}
\usepackage[T1]{fontenc}
\usepackage[latin9]{inputenc}
\usepackage[active]{srcltx}
\usepackage{amsmath}
\usepackage{graphicx}

\makeatletter

\providecommand{\tabularnewline}{\\}

\usepackage{aecompl}\usepackage{epstopdf}

\usepackage{babel}

\makeatother

\usepackage{babel}
\begin{document}
\title{Solute drag and dynamic phase transformations in moving grain boundaries}
\author{Y. Mishin}
\affiliation{Department of Physics and Astronomy, MSN 3F3, George Mason University,
Fairfax, Virginia 22030, USA}
\begin{abstract}
A discrete model and the regular solution approximation are applied
to describe the effect of grain boundary motion on grain boundary
phase transformations in a binary alloy. The model predicts all thermodynamic
properties of the grain boundary and the solute drag force, and permits
a broad exploration of the parameter space and different dynamic regimes.
The grain boundary phases continue to exist in the moving grain boundary
and show a dynamic hysteresis loop, a dynamic critical line, and other
features that are similar to those for equilibrium phases. Grain boundary
motion strongly affects the relative stability of the phases and can
even stabilize phases that are absolutely unstable under equilibrium
conditions. Grain boundary phase transformations are accompanied by
drastic changes in the boundary mobility. The results are analyzed
in the context of non-equilibrium thermodynamics. Unresolved problems
and future work are discussed.\bigskip{}
\end{abstract}
\email{Email address: ymishin@gmu.edu (Y. Mishin)}

\maketitle

\section{Introduction}

The phenomenon of solute drag, i.\,e., the resistance to grain boundary
motion caused by solute segregation, can strongly affect the mobility
of grain boundaries and thus microstructure development and properties
of alloys \citep{Balluffi95}. The first quantitative models of the
solute drag were proposed by Cahn \citep{Cahn1962} and Lücke et al.~\citep{Lucke-Stuwe-1963,Lucke:1971aa}.
One of the important predictions of these models is the highly nonlinear
relation between the solute drag force and the grain boundary speed,
with a maximum separating two different dynamic regimes. In the low-speed
regime, the boundary drags the segregation atmosphere with it, while
in the high-speed regime it breaks away from the old atmosphere and
forms a new one that poses less resistance to its motion. In recent
years, the solute drag effect was studied by the phase field \citep{Wang03,Li:2009aa,Shahandeh:2012aa,Gronhagen:2007aa,Abdeljawad:2017aa}
and phase-field crystal \citep{Greenwood:2012aa} methods, and by
atomic-level computer simulations \citep{Mendelev02a,Sun:2014aa,Wicaksono:2013aa,Mendelev:2001aa,Rahman:2016aa,Kim:2008aa}.

The simulation approaches mentioned above have their limitations.
For example, the phase-field models rely on the diffuse-interface
approximation and contain unknown parameters, such as the interface
mobility coefficients, which need to be calibrated using experimental
data. But it is exactly the interface mobility that needs to be predicted
as a function of the alloy composition and grain boundary speed. The
molecular dynamics (MD) method provides access to all atomic details
of the drag process, can be quantitatively accurate, and can describe
both thermodynamics and kinetics of the grain boundary motion using
the same model of atomic interactions. In an ideal world, this would
probably be the most effective tool for studying the solute drag.
Unfortunately, the speeds of today's computers strongly limit the
timescale accessible by MD simulations, making them too short for
a proper description of the solute drag. While diffusion along the
grain boundary can be modeled reasonably well \citep{Balluffi95,Mishin2010a,Mishin:2015ac},
diffusion in the surrounding lattice regions does not practically
occur on the MD timescale. In substitutional solid solutions, the
solute atoms diffuse by exchanges with lattice vacancies. Given the
small vacancy concentration and the relatively low vacancy jump rate,
in a typical MD simulation an average solute atom can only make a
few jumps at best.

Furthermore, most of the computer simulations, as well as the analytical
models \citep{Cahn1962,Lucke-Stuwe-1963,Lucke:1971aa}, have been
focused on dilute alloys for the sake of simplicity. Some of the phase-field
simulations do treat non-dilute solutions within the regular solution
approximation \citep{Abdeljawad:2017aa}, but the phase separation
is only considered as part of the bulk phase diagram. With rare exceptions
\citep{Wang03}, the effect of the segregation-induced grain boundary
phase transformations on the solute drag has not been analyzed. Meanwhile,
recent experiments suggest that grain boundaries are capable of first-order
phase transformations, in which their properties undergo discontinuous
changes \citep{Cantwell-2013,Baram08042011,Luo23092011,Harmer08042011,Rheinheimer201568}.
These phase transformations (sometimes called ``complexion transitions''
\citep{Cantwell-2013}) can affect many processes in materials. So
far, the theoretical models \citep{Cahn1962,Lucke-Stuwe-1963,Lucke:1971aa,Hart:1972aa,Cahn82a,Rottman1988a,Frolov:2015ab}
and computer simulations \citep{Frolov2013,Frolov2013a,Frolov:2015aa,Frolov:2016aa,Hickman__2017a,Yang:2018ab,Frolov:2018aa}
of the interface phase transformations have only dealt with systems
in thermodynamic equilibrium. Phase transformations in moving grain
boundaries have received little attention. This is unfortunate because
the transformations seen in experiments often occur in grain boundaries
that move due to the grain growth or other non-equilibrium processes.

In this paper we analyze the solute drag process within a discrete
model that describes the grain boundary thermodynamics, the grain
boundary motion, and the solute diffusion process within the same
theoretical framework. The model leaves many things out and only captures
the most important features of the solute drag process. The solute
diffusion is treated phenomenologically by relating the diffusion
flux to the chemical potential difference between crystal planes using
a composition and coordinate dependent diffusion mobility coefficient.
This enables us to overcome the timescale limitation of MD and perform
a parametric study of different dynamic regimes of the drag process.
Most importantly, the model predicts a miscibility gap in the grain
boundary, and thus the existence of two grain boundary phases with
different solute segregation levels. The model is applied to study
the effect of grain boundary motion on the grain boundary phases,
and conversely, the effect of the phase transformations on the grain
boundary mobility.

A similar model was previously developed by Wynblatt and Chatain \citep{Wynblatt2006,Wynblatt:2006aa,Wynblatt2008,Wynblatt:2008aa},
who applied it to describe segregation and phase transformations in
equilibrium grain boundaries. Their model was extended by Ma et al.~\citep{Wang03}
to moving grain boundaries and the results were compared with phase-field
simulations. We take the work by Ma et al.~\citep{Wang03} further
by making some improvements (such as the direct calculation of the
drag force and the entropy production rate) and performing a systematic
exploration of the parameter space, focusing on the effect of grain
boundary motion on the phase stability and phase transformations.
The results provide interesting insights into the solute drag phenomenon,
but also raise some general questions that are discussed in the end
of the paper.

\section{Model formulation}

\subsection{Thermodynamics and kinetics of the solid solution}

\begin{figure}
\includegraphics[width=0.7\textwidth]{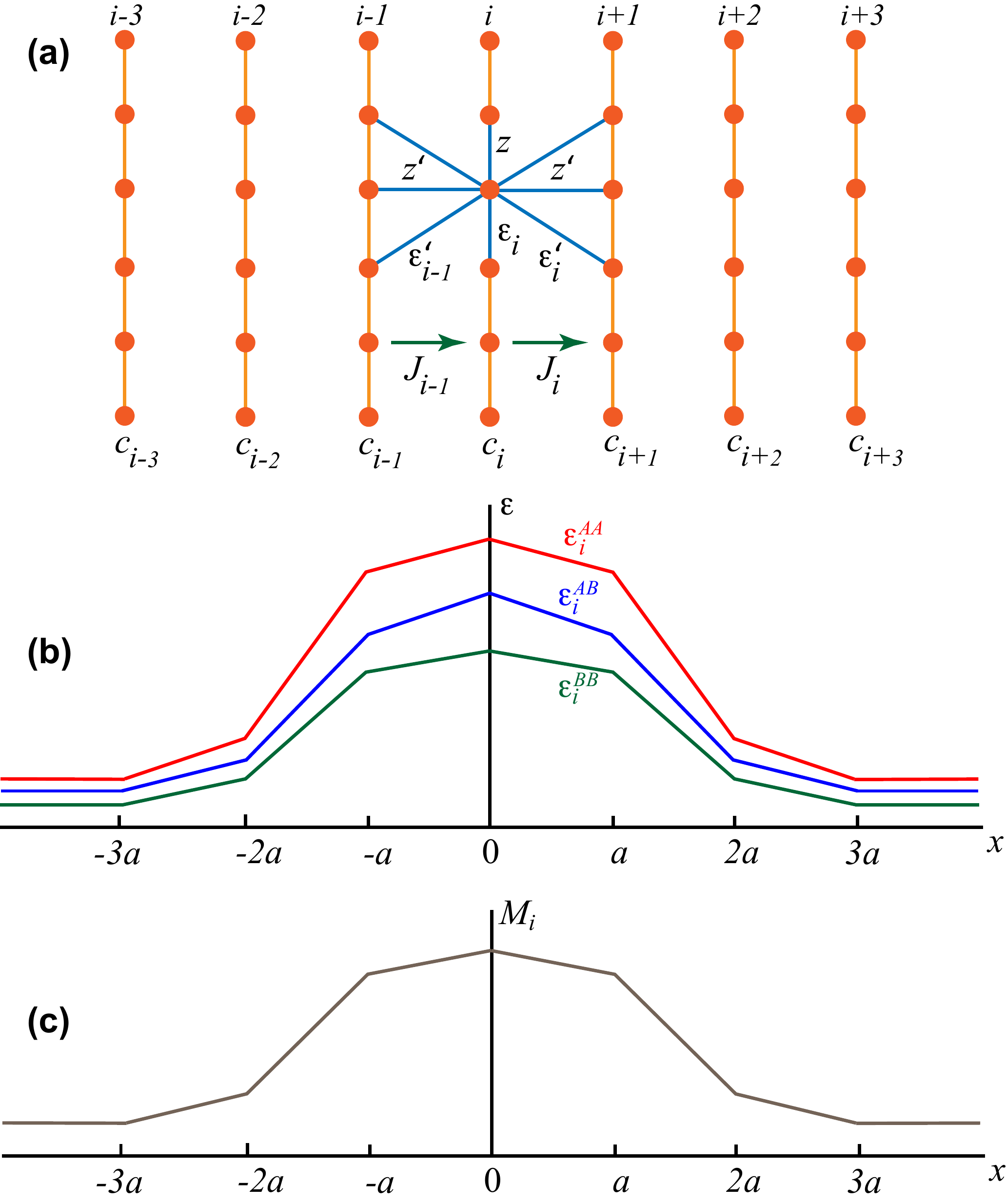}

\caption{(a) Schematic structure of a moving grain boundary, showing parallel
crystal planes numbered by index $i$ and separated by a spacing $a$.
(b) Coordinate dependence of the interaction parameters in the grain
boundary region (c) Coordinate dependence of the diffusion mobility
in the grain boundary region.\label{fig:schematic_model}}

\end{figure}

Following Wynblatt and Chatain \citep{Wynblatt2006,Wynblatt:2006aa,Wynblatt2008,Wynblatt:2008aa},
we consider a stack of identical crystal planes normal to the $x$
axis and separated by a spacing $a$ (Fig.~\ref{fig:schematic_model}(a)).
The planes are labeled by the index $i$ increasing in the positive
$x$ direction. Each atomic site has $z$ nearest neighbors in the
same plane $i$ and $z^{\prime}$ nearest neighbors in each of the
planes $(i-1)$ and $(i+1)$. Thus the total number of nearest neighbors
of a site is $z_{0}=z+2z^{\prime}$. 

The sites are populated by two sorts of atoms: A (solvent) and B (solute)
without vacancies. The chemical composition of any plane $i$ is characterized
by the fraction $c_{i}$ of atoms B. It is assumed that the atomic
interactions are pairwise, limited to nearest neighbors, and follow
the regular solution model. The interaction parameters are denoted
$\varepsilon_{i}^{AA}$, $\varepsilon_{i}^{BB}$ and $\varepsilon_{i}^{AB}$
for interatomic bonds within each layer $i$, and $\varepsilon_{i}^{\prime AA}$,
$\varepsilon_{i}^{\prime BB}$ and $\varepsilon_{i}^{\prime AB}$
for bonds between neighboring planes $i$ and $(i+1)$. 

Any distribution of the solute atoms over the system is fully described
by a discrete set of layer compositions $\left\{ c_{i}\right\} $.
The total free energy of the system per unit cross-sectional area
is 
\begin{eqnarray}
F & = & \dfrac{z}{2s}\sum_{i}\left[\varepsilon_{i}^{AA}+(\varepsilon_{i}^{BB}-\varepsilon_{i}^{AA})c_{i}+\omega_{i}c_{i}(1-c_{i})\right]\nonumber \\
 & + & \dfrac{z^{\prime}}{s}\sum_{i}\left[\varepsilon_{i}^{\prime AA}+(\varepsilon_{i}^{\prime BB}-\varepsilon_{i}^{\prime AA})\dfrac{c_{i}+c_{i+1}}{2}+\omega_{i}^{\prime}\left(\dfrac{c_{i}+c_{i+1}}{2}-c_{i}c_{i+1}\right)\right]\nonumber \\
 & + & \dfrac{kT}{s}\sum_{i}\left[c_{i}\ln c_{i}+(1-c_{i})\ln(1-c_{i})\right].\label{eq:F}
\end{eqnarray}
Here $s$ is the area per site, $k$ is Boltzmann's constant, $T$
is temperature, and 
\[
\omega_{i}=2\varepsilon_{i}^{AB}-\varepsilon_{i}^{AA}-\varepsilon_{i}^{BB}
\]
and
\[
\omega_{i}^{\prime}=2\varepsilon_{i}^{\prime AB}-\varepsilon_{i}^{\prime AA}-\varepsilon_{i}^{\prime BB}
\]
are the intra-layer and inter-layer regular solution parameters, respectively.
The last term in Eq.(\ref{eq:F}) comes from the ideal entropy of
mixing in the mean-field approximation. We can also compute the derivatives
\begin{eqnarray}
\mu_{i}=s\dfrac{\partial F}{\partial c_{i}} & = & z\left(\varepsilon_{i}^{AB}-\varepsilon_{i}^{AA}\right)+z^{\prime}\left(\varepsilon_{i}^{\prime AB}-\varepsilon_{i}^{\prime AA}+\varepsilon_{i-1}^{\prime AB}-\varepsilon_{i-1}^{\prime AA}\right)\nonumber \\
 & - & z\omega_{i}c_{i}-z^{\prime}\omega_{i}^{\prime}c_{i+1}-z^{\prime}\omega_{i-1}^{\prime}c_{i-1}+kT\ln\dfrac{c_{i}}{1-c_{i}}.\label{eq:mu}
\end{eqnarray}
The quantity $\mu_{i}$ has the meaning of the diffusion potential
\citep{Larche_Cahn_78,Larche1985} of atoms B relative to atoms A
in plane $i$. For brevity, we will be referring to $\mu_{i}$ as
the chemical potential of the solute atoms.

In the particular case of a homogeneous solution representing interior
regions of the grains, the free energy per atom is
\begin{eqnarray}
\varphi & = & \dfrac{1}{2}z_{0}\varepsilon^{AA}+\dfrac{1}{2}z_{0}\left(\varepsilon^{BB}-\varepsilon^{AA}\right)c+\dfrac{1}{2}z_{0}\omega c(1-c)\nonumber \\
 & + & kT\left[c\ln c+(1-c)\ln(1-c)\right]\label{eq:f_homo}
\end{eqnarray}
while the chemical potential is
\begin{equation}
\mu=\dfrac{1}{2}z_{0}\left(\varepsilon^{BB}-\varepsilon^{AA}\right)+\dfrac{1}{2}z_{0}\omega(1-2c)+kT\ln\dfrac{c}{1-c},\label{eq:mu_homo}
\end{equation}
where $\omega=2\varepsilon^{AB}-\varepsilon^{AA}+\varepsilon^{BB}$
is the regular solution parameter inside the grains and $c$ is the
grain composition. The intra-layer and inter-layer interaction parameters
inside the grains are assumed to be identical, making the grain properties
isotropic. If $\omega>0$, the homogeneous solution separates in two
solid-solution phases below the critical temperature $T_{c}=z_{0}\omega/4k$.
The miscibility gap on the $T$-$c$ phase diagram  is symmetric and
is described by the equation
\[
T=\dfrac{z_{0}\omega(c-1/2)}{k\ln\dfrac{c}{1-c}}
\]
for the phase coexistence line (binodal) and
\[
T=\dfrac{z_{0}\omega c}{k}(1-c)
\]
for the spinodal line. 

We further assume that the solute atoms can diffuse through the system
by hopping between neighboring crystal planes. The diffusive flux
$J_{i}$ from plane $i$ to plane $(i+1)$ is driven by the chemical
potential difference between the planes according to the equation
\begin{equation}
J_{i}=-M_{i}c_{i}(1-c_{i+1})(\mu_{i+1}-\mu_{i}),\label{eq:flux}
\end{equation}
$M_{i}$ being the diffusion mobility coefficient. For a given composition
profile $\left\{ c_{i}\right\} $, Eq.(\ref{eq:mu}) predicts the
chemical potentials $\left\{ \mu_{i}\right\} $ in the planes and
Eq.(\ref{eq:flux}) gives the diffusion fluxes $\left\{ J_{i}\right\} $
between the planes. The rate of compositional change $\dot{c}_{i}$
in every plane $i$ is then obtained from the continuity equation
\begin{equation}
\dot{c}_{i}=s(J_{i-1}-J_{i}).\label{eq:continuity}
\end{equation}
This equation, in conjunction with Eqs.(\ref{eq:mu}) and (\ref{eq:flux}),
can be solved numerically by a finite-difference method to predict
the time evolution of the composition profile.

\subsection{The grain boundary model}

The grain boundary is represented by a region in which the thermodynamic
and kinetic parameters are different from those outside this region.
Specifically, we will follow Ma et al.~\citep{Wang03} to represent
the interaction parameters by the Gaussian functions
\begin{equation}
\varepsilon_{i}^{\nu}=\hat{\varepsilon}^{\nu}\exp\left(-\dfrac{(x-ai)^{2}}{\lambda^{2}}\right)+\varepsilon^{\nu},\label{eq:Gauss_1}
\end{equation}
\begin{equation}
\varepsilon_{i}^{\prime\nu}=\hat{\varepsilon}^{\prime\nu}\exp\left(-\dfrac{(x-a/2-ai)^{2}}{\lambda^{2}}\right)+\varepsilon^{\nu}.\label{eq:Gauss_2}
\end{equation}
Similarly, for the diffusion mobility
\begin{equation}
M_{i}=\hat{M}\exp\left(-\dfrac{(x-a/2-ai)^{2}}{\lambda^{2}}\right)+M.\label{eq:Gauss_3}
\end{equation}
In these equations, $x$ and $\lambda$ are the Gaussian position
and width, respectively. The chemical symbol $\nu$ represents the
pairs AA, BB and AB. $M$ is the diffusion mobility inside the grains.
The symbols with the hat represent the maxima of the respective properties
in the grain boundary region. The Gaussians (\ref{eq:Gauss_1})-(\ref{eq:Gauss_3})
are illustrated schematically in Fig.~\ref{fig:schematic_model}(b,c)
for $x=0$. It should be noted that the Gaussian functions are only
chosen here for the sake of simplicity. Any other bell-shape function
would be suitable as well. 

The Gaussians (\ref{eq:Gauss_1}) and (\ref{eq:Gauss_2}) modify thermodynamic
properties of the alloy in the grain boundary region relative to the
grain interiors, creating a driving force for solute segregation.
Equation (\ref{eq:Gauss_3}) accounts for possible deviations (usually,
enhancement) of the solute diffusivity in the boundary region relative
to the grains. Note that the grain boundary is not endowed with any
distinct atomic structure. It is only different from the grains in
thermodynamic and kinetic properties. The grain boundary position
is identified with the coordinate $x$ of the Gaussians (\ref{eq:Gauss_1})-(\ref{eq:Gauss_3}).
The boundary can be moved with any desired speed $V=dx/dt$ by simply
shifting the Gaussians along the $x$-axis while keeping the crystal
planes fixed. In this process, the solute diffusion tries to catch
up with the boundary motion to maintain the segregation atmosphere.
This creates a competition between the two kinetic processes: the
grain boundary motion and the solute diffusion. Different outcomes
of this competition give rise to different regimes of the solute drag
process as will be discussed below.

\subsection{The solute drag force\label{subsec:The-solute-drag}}

To calculate the solute drag force, we will treat the system evolution
from the perspective of non-equilibrium thermodynamics, as was done
previously by Hillert et al.~\citep{Hillert:1976aa,Hillert:1999aa,Hillert:2001aa}.
In the present model, the system is assumed to be connected to a reservoir
of atoms maintained at a fixed chemical potential $\mu$ and temperature
$T$. Under such conditions, the rate of total free energy change
of the composite system (our system plus the reservoir) per unit grain
boundary area is
\begin{equation}
\dot{F}_{\textnormal{tot}}=\dot{F}-\mu\sum_{i}\dfrac{\dot{c}_{i}}{s},\label{eq:Dissipation_0}
\end{equation}
where the first term is the rate of free energy change in our system
(the grain boundary plus the grains) while the second term represents
the rate of free energy change of the reservoir due to the atomic
exchange with our system. Assuming that the composite system is closed
and remains in thermal equilibrium, $\dot{F}_{\textnormal{tot}}$
is expected to have the structure \citep{Prigogine1968,De-Groot1984}
\begin{equation}
\dot{F}_{\textnormal{tot}}=-T\sigma+\dot{W},\label{eq:Dissipation_1}
\end{equation}
where $\sigma$ is the irreversible entropy production rate and $\dot{W}$
is the work done on the system per unit time (input power).

The free energy $F$ is given by Eq.~(\ref{eq:F}). It varies with
time through the time-dependent concentrations $c_{i}$ and the $x$-dependent
interaction parameters $\varepsilon_{i}^{\nu}$ and $\varepsilon_{i}^{\prime\nu}$,
which change in time due to the boundary motion. Thus, the time derivative
$\dot{F}$ contains terms proportional to $\dot{c}_{i}$ and a term
proportional to the boundary speed $V$. Calculations show that
\begin{equation}
\dot{F}_{\textnormal{tot}}=\sum_{i}\left(\mu_{i}-\mu\right)\dfrac{\dot{c}_{i}}{s}+fV,\label{eq:Dissipation_1.1}
\end{equation}
where 
\begin{eqnarray}
f & = & -\sum_{i}\dfrac{z(x-ai)}{s\lambda^{2}}\left[\varepsilon_{i}^{AA}+(\varepsilon_{i}^{BB}-\varepsilon_{i}^{AA})c_{i}+\omega_{i}c_{i}(1-c_{i})\right]\nonumber \\
 & - & \sum_{i}\dfrac{2z^{\prime}(x-a/2-ai)}{s\lambda^{2}}\left[\varepsilon_{i}^{\prime AA}+(\varepsilon_{i}^{\prime BB}-\varepsilon_{i}^{\prime AA})\dfrac{c_{i}+c_{i+1}}{2}\right]\nonumber \\
 & - & \sum_{i}\dfrac{2z^{\prime}(x-a/2-ai)}{s\lambda^{2}}\omega_{i}^{\prime}\left(\dfrac{c_{i}+c_{i+1}}{2}-c_{i}c_{i+1}\right)\nonumber \\
 & + & \sum_{i}\dfrac{z(x-ai)}{s\lambda^{2}}\left[\varepsilon^{AA}+(\varepsilon^{BB}-\varepsilon^{AA})c_{i}+\omega c_{i}(1-c_{i})\right]\nonumber \\
 & + & \sum_{i}\dfrac{2z^{\prime}(x-a/2-ai)}{s\lambda^{2}}\left[\varepsilon^{AA}+(\varepsilon^{BB}-\varepsilon^{AA})\dfrac{c_{i}+c_{i+1}}{2}\right]\nonumber \\
 & + & \sum_{i}\dfrac{2z^{\prime}(x-a/2-ai)}{s\lambda^{2}}\omega\left(\dfrac{c_{i}+c_{i+1}}{2}-c_{i}c_{i+1}\right).\label{eq:f}
\end{eqnarray}
Substituting $\dot{c}_{i}$ from Eq.~(\ref{eq:continuity}) and applying
the zero-flux condition far away from the grain boundary, we finally
obtain
\begin{equation}
\dot{F}_{\textnormal{tot}}=\sum_{i}J_{i}(\mu_{i+1}-\mu_{i})+fV.\label{eq:Dissipation_2}
\end{equation}
The first term in this equation represents the free energy dissipation
rate $-T\sigma$ due to the diffusion process. Thus, the entropy production
rate is
\begin{equation}
\sigma=-\dfrac{1}{T}\sum_{i}J_{i}(\mu_{i+1}-\mu_{i}).\label{eq:entropy_production}
\end{equation}
The form of this term confirms that the diffusive flux $J_{i}$ is
driven by the chemical potential difference between neighboring planes,
as was assumed in Eq.(\ref{eq:flux}). The second term in Eq.~(\ref{eq:Dissipation_2})
is the input power $\dot{W}$ required for moving the grain boundary
with the speed $V$. Accordingly, the coefficient $f$ has the meaning
of the force (per unit area) driving the boundary motion.

\begin{figure}
\noindent \begin{centering}
(a)\enskip{}\includegraphics[width=0.65\textwidth]{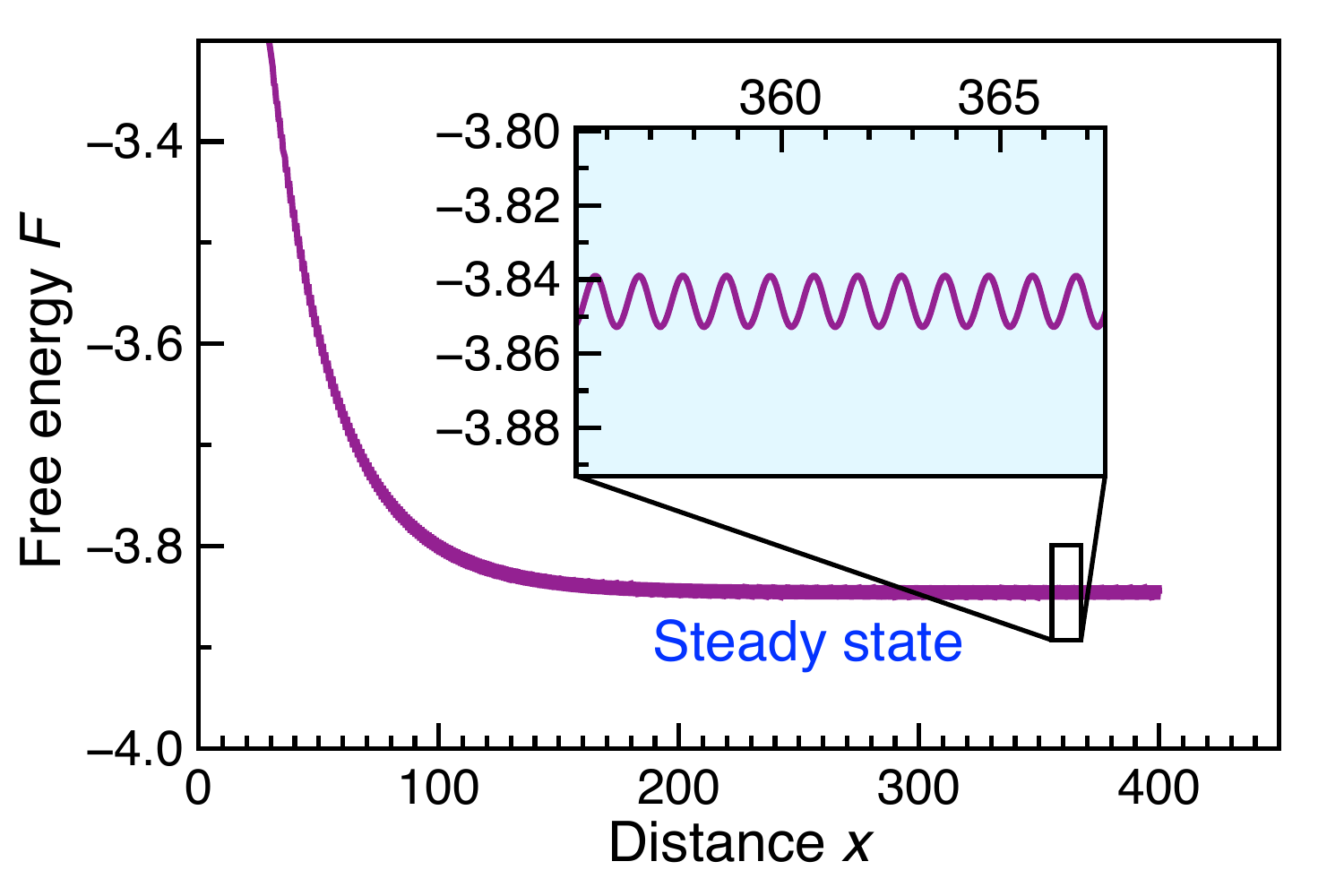}
\par\end{centering}
\noindent \begin{centering}
(b)\enskip{}\includegraphics[width=0.65\textwidth]{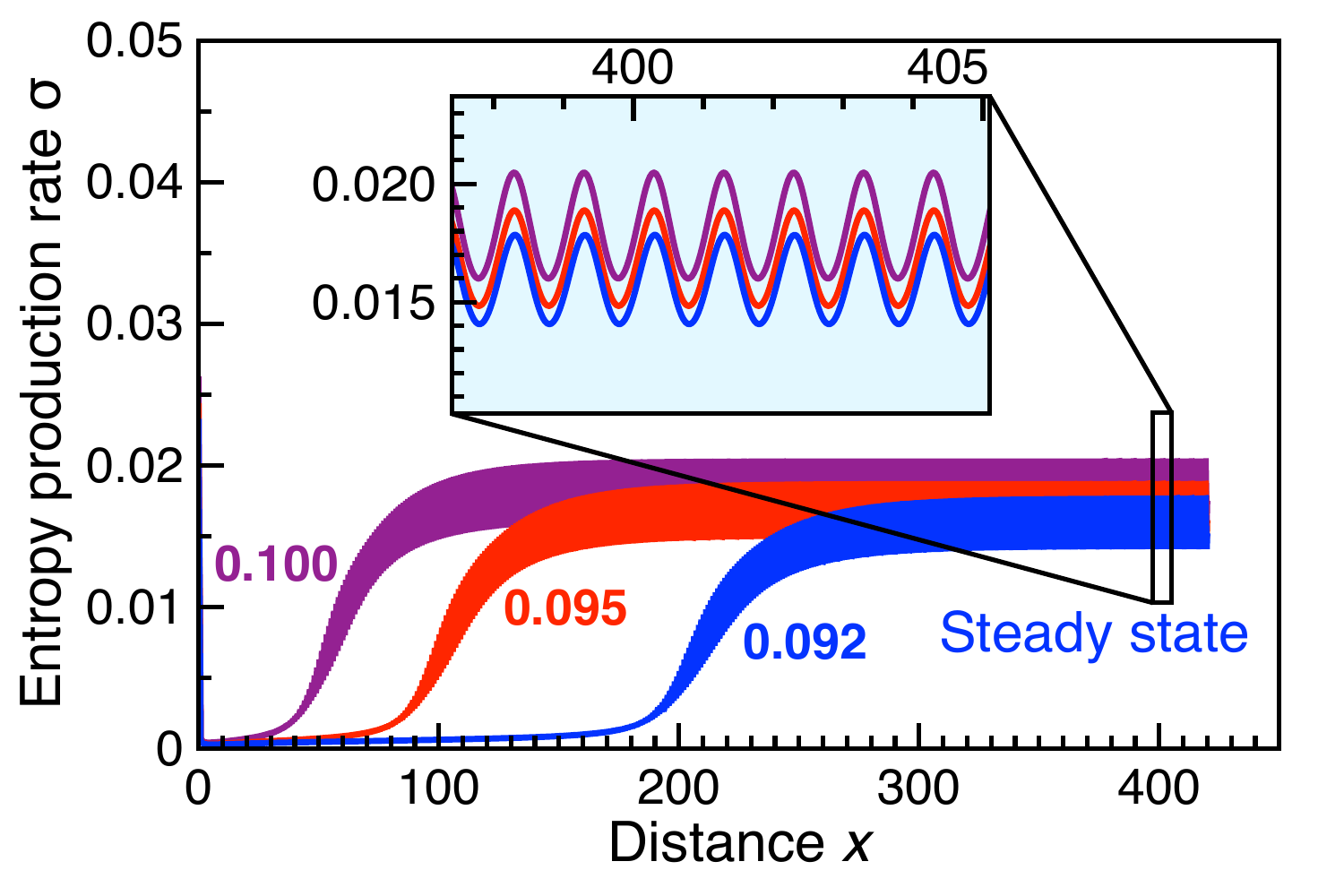}
\par\end{centering}
\caption{(a) Typical plot of the free energy $F$ as a function of the distance
$x$ travelled by the grain boundary with the speed $V=0.05$ starting
from a uniform alloy. The temperature is $T=0.8$ and the alloy composition
is $c=0.1$. (b) Same for the entropy production rate $\sigma$ and
the alloys compositions $c=0.092$, 0.095 and 0.1. The inset shows
the oscillations with the period 1 arising due to the discrete character
of the model. All variables are normalized as indicated in Table \ref{tab:Dimensionless-parameters}.\label{fig:Oscillations}}

\end{figure}

We are especially interested in the case of steady state boundary
motion. We first need to clarify the meaning of steady state in this
model. Due to the discrete character of the model, the interaction
parameters on either side of the moving Gaussian peak slightly vary
as the peak traverses the spaces between neighboring atomic planes.
This causes slight oscillations of the free energy, the driving force,
the entropy production rate, and all grain boundary properties. Such
oscillations are illustrated in Fig.~\ref{fig:Oscillations} where
the free energy of the system $F$ and the entropy production rate
$\sigma$ are plotted as functions of the distance travelled by the
grain boundary in a typical computer simulation run. We use the normalized
variables summarized in Table \ref{tab:Dimensionless-parameters}.
The simulation details will be explained later; at this point we only
want to demonstrate the oscillations. Fig.~\ref{fig:Oscillations}(a)
shows that, after the initial decrease, the free energy reaches a
plateau in which it continues to oscillate with a constant amplitude
around a constant value. The same behavior is exhibited by $\sigma$
(Fig.~\ref{fig:Oscillations}(b)), and by all other properties of
the system in the long-time limit. We interpret this plateau as a
steady state despite the existence of oscillations. Since the oscillations
per se do not present interest in this work, all results reported
below have been averaged over a period of oscillations. In this average
sense, the steady state motion does satisfy the standard condition
$\dot{F}_{\textnormal{tot}}=0$. In the steady state regime, the input
power $\dot{W}$ is fully dissipated by diffusion, keeping the period-averaged
value of the total free energy constant. The driving force $f$ is
then equal to the solute drag (friction) force resisting the boundary
motion.

It should be noted that in the present model, the resistance to the
grain boundary motion is purely chemical. In a single-component system
($c=0$ or $c=1$), the force (\ref{eq:f}) averaged over the oscillations
is zero. Thus the boundary motion becomes frictionless, which is of
course not realistic. A nonzero intrinsic mobility could be introduced
by adding another, composition-independent term in Eq.~(\ref{eq:f}).
We choose not to do so to focus the attention solely on the solute
drag effect.

\subsection{The grain boundary segregation and free energy}

When the grain boundary is stationary ($V=0$), it eventually reaches
thermodynamic equilibrium with the grains, forming an equilibrium
segregation atmosphere. The amount of segregation can be characterized
by 
\begin{equation}
\Gamma=\dfrac{1}{s}\sum_{i}(c_{i}-c).\label{eq:segreg}
\end{equation}

The grain boundary free energy $\gamma$ is defined as the reversible
work expended on the formation of a unit boundary area in a closed
system. This definition is equivalent to the excess (per unit area)
of the grand potential $F-\mu N$, $N$ being the total number of
the solute atoms in a sample of unit-area. Thus,
\begin{equation}
\gamma=F-F_{\textnormal{hom}}-\mu\Gamma,\label{eq:gamma}
\end{equation}
where $F_{\textnormal{hom}}=(1/s)\sum_{i}\varphi$ is the free energy
of the homogeneous state of the system in the absence of the grain
boundary. In this equation, $F$ and $\varphi$ are given by Eqs.~(\ref{eq:F})
and (\ref{eq:f_homo}), respectively.

Formally, Eq.~(\ref{eq:gamma}) can also be applied to non-equilibrium
states of the system as long as the atmosphere is localized near the
grain boundary so that the sums over $i$ converge. While the usefulness
of this $\gamma$ in general is questionable, it becomes a useful
grain boundary property in the case of steady state motion as will
be discussed later.

\section{Numerical calculations}

\subsection{Dimensionless parameters}

For the numerical calculations presented below, the model parameters
were transformed to dimensionless forms as specified in Table \ref{tab:Dimensionless-parameters}.
The temperature was normalized by the bulk critical point $T_{c}$
and all energies by $kT_{c}$. The units of length and time become
$a$ and $Dt/a^{2}$, respectively. Thus, the solute atoms diffuse
over a distance comparable to $a$ per unit time. The dimensionless
equations of the model are obtained from the physical equations presented
above by simply dropping the coefficients $a$, $s$ and $k$. From
this point on, all calculations will be discussed in terms of these
dimensionless variables.

The dimensionless grain boundary speed $aV/D$ is a measure of the
relative importance of the two kinetic processes: the grain boundary
motion and the solute diffusion. When this dimensionless speed is
$\gg1$, the boundary motion dominates over diffusion; when it is
$\ll1$, diffusion is the fastest process. In the first case, we can
expect that the solute diffusion will not be able to catch up with
the grain boundary motion and the segregation atmosphere formed at
$V=0$ will be left behind the moving boundary. In the second case,
diffusion can keep up with the boundary motion and the atmosphere
can be dragged along with the boundary. 

\subsection{Grain boundary thermodynamics\label{subsec:GB-thermodynamics}}

Before analyzing the solute drag process, we will examine the equilibrium
thermodynamic properties of the grain boundary predicted by this model.
The segregation atmosphere can be computed from the equilibrium condition
$\mu_{i}=\mu$ for all $i$, with $\mu_{i}$ and $\mu$ given by Eqs.~(\ref{eq:mu})
and (\ref{eq:mu_homo}), respectively. This condition generates a
set of recursive equations that can be solved numerically \citep{Wynblatt2006,Wynblatt:2006aa,Wynblatt2008,Wynblatt:2008aa}.
Here we used a different methods. Namely, the diffusion equations
(\ref{eq:flux}) and (\ref{eq:continuity}) were solved numerically
until all fluxes became negligibly small ($\sim10^{-10}$), signifying
equilibrium. Multiple initial conditions were tested to identify all
stable or metastable states. The diffusion equations were evolved
using an explicit-in-time finite-difference scheme. Typical results
will be illustrated below for the following set of dimensionless parameters:
$z=6$, $z^{\prime}=3$, $\lambda=1.4$, $\varepsilon^{AA}=\varepsilon^{BB}=0$,
$\varepsilon^{AB}=0.167$, $\hat{\varepsilon}^{AA}=\hat{\varepsilon}^{\prime AA}=0.35$,
$\hat{\varepsilon}^{BB}=\hat{\varepsilon}^{\prime BB}=0.20$, $\hat{\varepsilon}^{AB}=\hat{\varepsilon}^{\prime AB}=0.325$,
$M=1$ and $\hat{M}=2$.

\begin{figure}
\begin{centering}
\textbf{(a)}\enskip{}\includegraphics[width=0.5\textwidth]{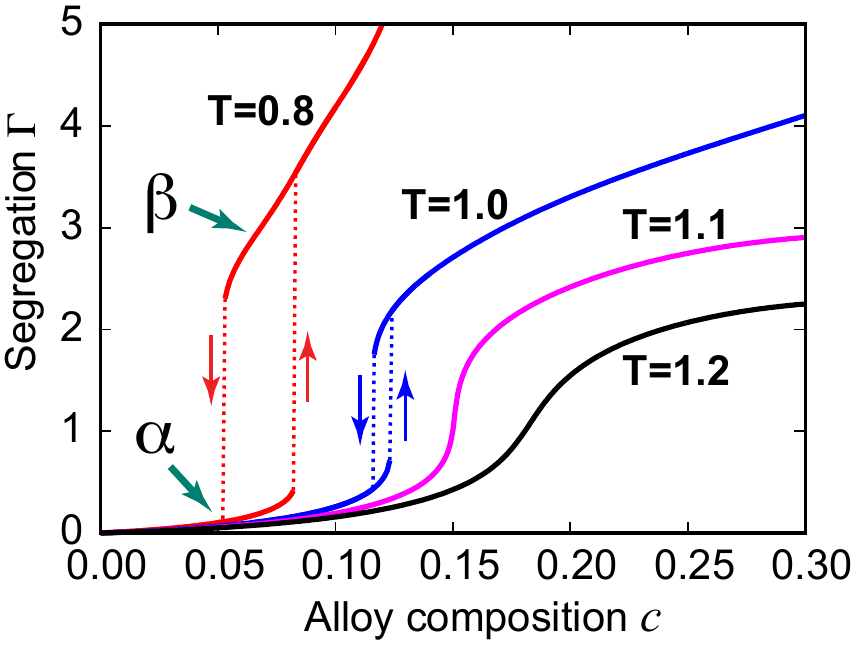}
\par\end{centering}
\bigskip{}

\begin{centering}
\textbf{(b)}\enskip{}\includegraphics[width=0.52\textwidth]{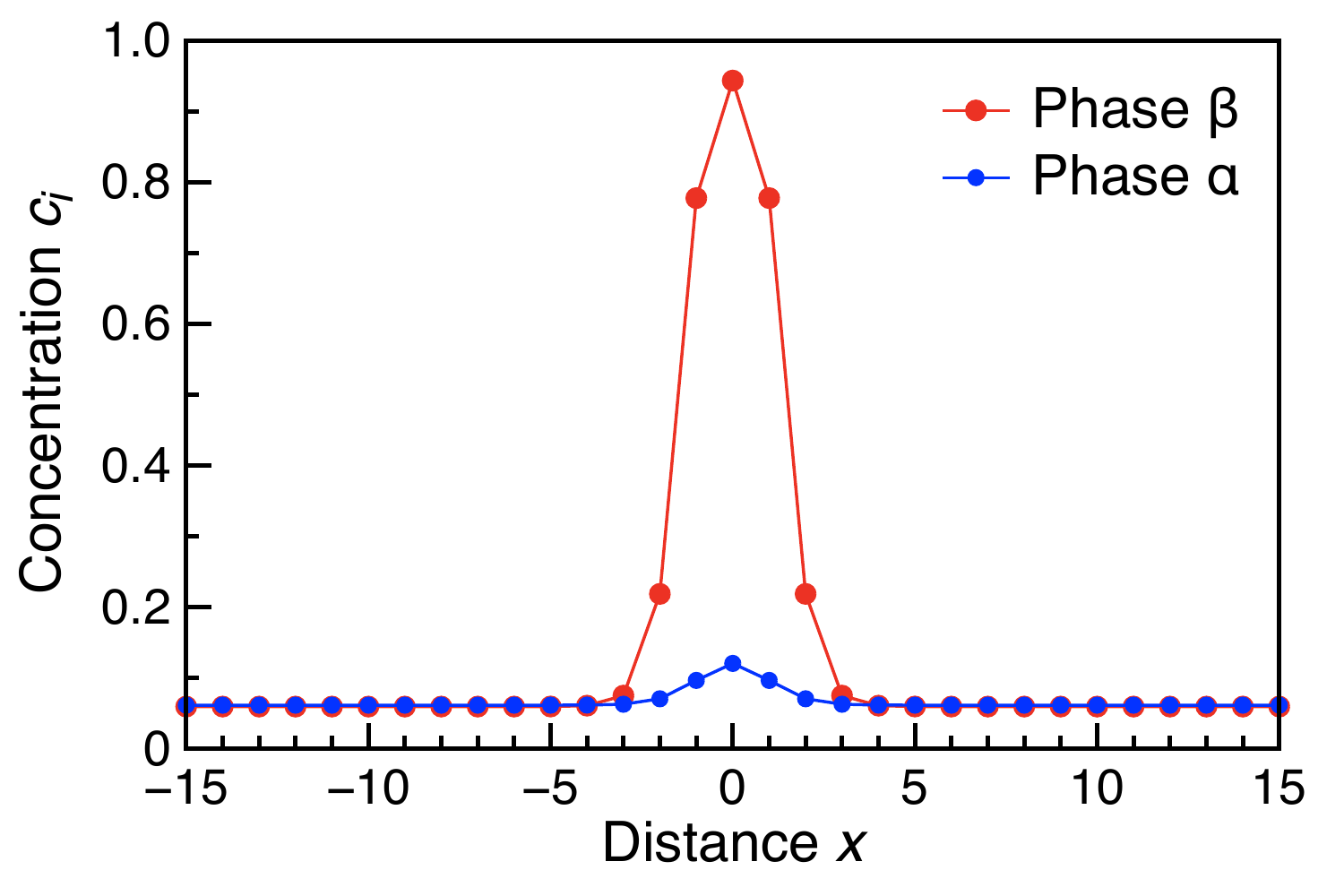}
\par\end{centering}
\caption{(a) Equilibrium segregation isotherms at selected temperatures showing
the discontinuities and hysteresis caused by transformations between
grain boundary phases $\alpha$ (low segregation) and $\beta$ (high
segregation). (b) Equilibrium segregation profiles for the grain boundary
phases coexisting at the temperature $T=0.8$ and grain composition
$c=0.063$.\label{fig:Equi_1}}

\end{figure}

As expected from the regular solution model, phase separation occurs
inside the grain boundary region. Figure \ref{fig:Equi_1}(a) illustrates
solute segregation isotherms $\Gamma(c)$ at several temperatures.
At each temperature, the grain composition $c$ was varied over an
interval in which the grains remained in a single-phase state. At
relatively low temperatures, the isotherms display a discontinuity
and a hysteresis. This behavior identifies two grain boundary phases:
a low-segregation phase $\alpha$ and a high-segregation phase $\beta$
(Fig.~\ref{fig:Equi_1}(b)). If the grain composition is varied back
and forth across the hysteresis region, the height of the concentration
peak in the grain boundary region jumps up and down as the boundary
switches abruptly from one phase to the other. As temperature increases,
the miscibility gap between the grain boundary phases narrows and
eventually closes at a critical point $T_{GB}$. For this particular
choice of the model parameters, the grain boundary critical point
$T_{GB}=1.072$ lies slightly above the bulk critical point $T_{c}=1$.

Figure \ref{fig:gamma_plot} shows the isotherms of the grain boundary
free energy $\gamma$. Above $T_{GB}$, $\gamma$ is a smooth nonlinear
function of the bulk composition all the way from pure component A
to pure B. For the chosen set of model parameters, the single-component
B system has a lower grain boundary free energy than the single-component
A system. Below $T_{GB}$, the curves remain continuous but develop
a discontinuous first derivative. This is best seen in the inset,
where the curves represent the free energies of the two grain boundary
phases at $T=0.8$. The crossing point of the curves corresponds to
the grain boundary phase coexistence state. On the right-hand side
of the coexistence point, the $\beta$ phase becomes thermodynamically
more stable than the $\alpha$ phase. However, both curves continue
past the coexistence point due to the hysteresis effect, giving rise
to metastable branches that terminate at two spinodal points (open
circles). 

The results of the calculations are summarized in Fig.~\ref{fig:Phase_diag}
where the grain boundary phase transformation lines are superimposed
on the bulk phase diagram. Similar to the bulk phase transformations,
the grain boundary transformations exhibit a phase coexistence line
and two spinodal lines, which merge at the grain boundary critical
point. By adjusting the model parameters, the grain boundary critical
point can be placed anywhere above the bulk miscibility gap. It should
be noted that the diagram does not show the miscibility gap between
the grain boundary phases. The reason is that the grain boundary is
treated as an open system connected to the grains playing the role
of an infinitely large reservoir imposing fixed values of $\mu$ and
$T$. In this grand-canonical treatment, the total amount of solute
in the grain boundary region cannot be controlled. It is automatically
adjusted to achieve thermodynamic equilibrium in the system. As $T$
and $\mu$ vary, the boundary switches from one single-phase state
to another without having to partition the solute between two phases.
A more detailed discussion of interface phase transformations and
interface phase rules can be found elsewhere \citep{Frolov:2015ab}.

The phase diagram in Fig.~\ref{fig:Phase_diag} permits predictions
of possible scenaria of the grain boundary phase changes. For example,
if the grain boundary is initially in the low-segregation phase $\alpha$
and the sample is cooled down, then at some point the segregation
must abruptly increase as the boundary transforms to phase $\beta$.
Alternatively, we can fix the temperature and keep adding more solute
to the sample, until the segregation jumps to a higher level when
the grain boundary transforms to phase $\beta$. In both cases, the
grain boundary free energy $\gamma$ remains continuous if the transformation
occurs under equilibrium conditions, or decreases discontinuously
if the $\alpha$ phase gets undercooled or oversaturated.

\subsection{The solute drag effect}

Grain boundary motion was modeled by imposing a constant grain boundary
speed $V$ at fixed values of temperature and chemical potential (and
thus the bulk composition). The diffusion equations (\ref{eq:flux})
and (\ref{eq:continuity}) were evolved until the system reached a
steady state, in which the composition profile, the total free energy
and the drag force (averaged over the period of oscillations, see
Section \ref{subsec:The-solute-drag} and Fig.~\ref{fig:Oscillations})
remained constant within numerical accuracy. Some of the simulations
were started from one of the equilibrium grain boundary phases and
the speed was increased by small increments. The steady state composition
profile obtained at each step was used as the initial condition for
the next increment of speed. In other cases, the speed was decreased
by small increments until the grain boundary motion stopped. In yet
another type of simulations, a chosen grain boundary speed was instantly
applied to an equilibrium grain boundary. In the latter case, the
goal was to determine if the segregation atmosphere can follow the
grain boundary motion. A systematic parametric study was performed,
exploring the space of the three variables $T$, $c$ and $V$. Representative
results will be presented below for the particular set of model parameters
indicated in Section \ref{subsec:GB-thermodynamics}.

A typical plot of the drag force as a function of grain boundary speed
is shown in Fig.~\ref{fig:f-V_1}. In this example, the alloy composition
and temperature are represented by point A on the phase diagram in
Fig.~\ref{fig:Phase_diag}(b). The boundary was initially in the
equilibrium state corresponding to phase $\beta$. It was then brought
to motion by gradually increasing the speed. The force-speed relation
obtained is linear at $V\ll1$ and develops downward deviations from
linearity and eventually turns over as the speed increases. The two
branches of the curve correspond to different dynamic regimes. On
the left-hand (rising) branch, the resistance to the boundary motion
increases with the speed. To move the boundary faster, a larger force
must be applied. On the right-hand (falling) branch, the force decreases
with speed. The faster the boundary moves, the easier it is to move
it. As will be discussed later, this may result in a morphological
instability of the grain boundary under certain conditions.

\begin{figure}
\noindent \begin{centering}
\includegraphics[width=0.6\textwidth]{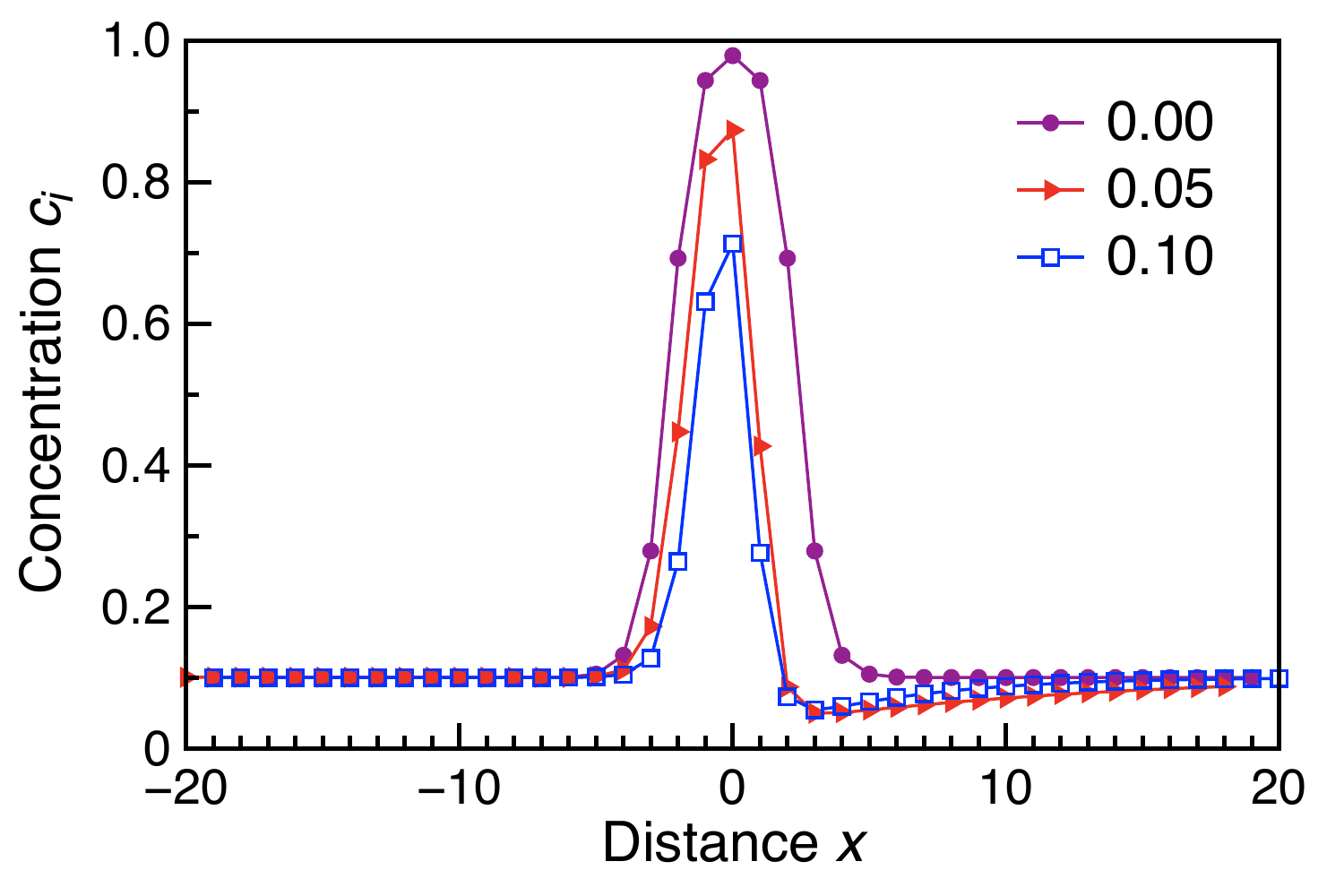}
\par\end{centering}
\caption{Grain boundary composition profiles for the speeds $V$ indicated
in the key. The reference frame is attached to the moving grain boundary.
The alloy composition and temperature are $c=0.1$ and $T=0.8$, respectively.\label{fig:GB-profiles-1}}

\end{figure}

The shape of the dynamic segregation profile varies continuously as
the speed increases (until some point that will be discussed later).
The motion breaks the symmetry of the profile, creating a depletion
zone ahead of the boundary and reducing the height of the peak. The
chemical composition behind the moving boundary remains very close
to the bulk composition. The evolution of the segregation atmosphere
with the increasing speed is illustrated in Fig.~\ref{fig:GB-profiles-1},
where the composition profiles are plotted in the reference frame
moving with the boundary. The profile shapes are qualitatively consistent
with predictions of Cahn's model \citep{Cahn1962}. As shown in Fig.~\ref{fig:Properties_A}(a),
the total amount of grain boundary segregation decreases with the
boundary speed; the boundary gradually loses part of its atmosphere.
At the same time, the dynamic grain boundary free energy $\gamma$
computed from Eq.~(\ref{eq:gamma}) increases with the speed (Fig.~\ref{fig:Properties_A}(b)).
This creates an inverse correlation between $\gamma$ and the amount
of segregation, which is consistent with the similar trend in equilibrium
interface thermodynamics \citep{Willard_Gibbs}.

\begin{figure}
\includegraphics[width=0.45\columnwidth]{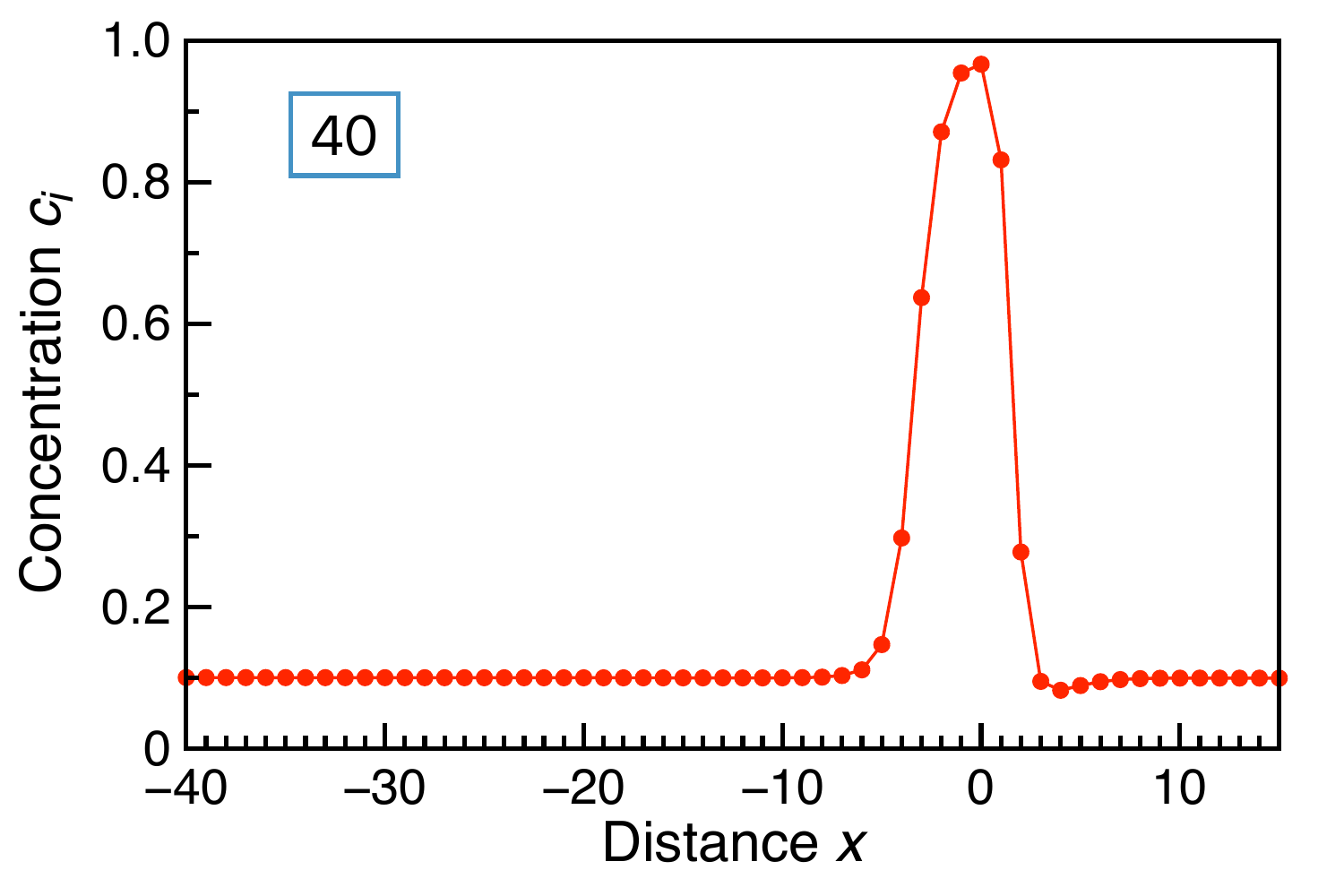}\quad{}\includegraphics[width=0.45\columnwidth]{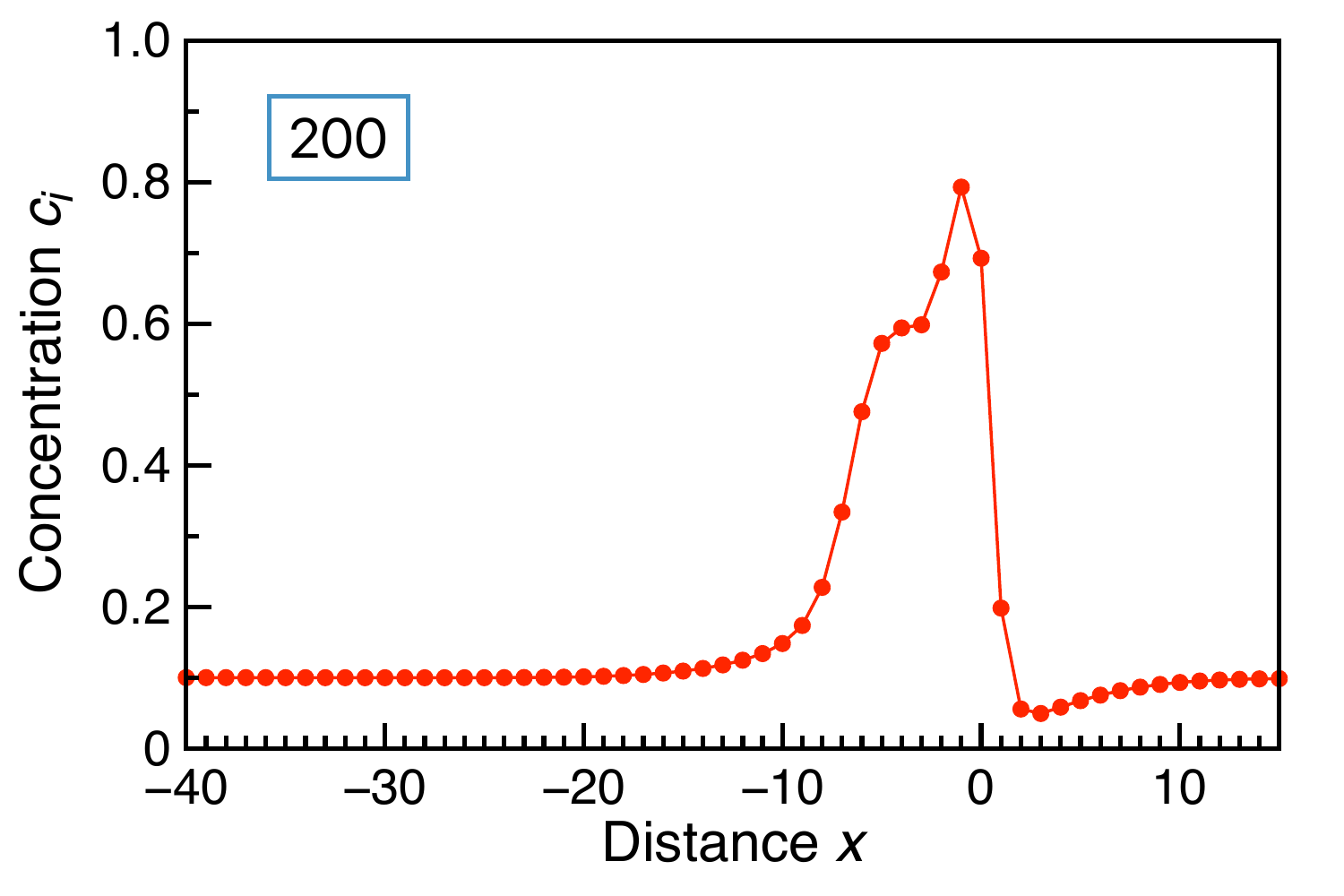}

\textbf{(a)}\quad{}\quad{}\quad{}\quad{}\quad{}\quad{}\quad{}\quad{}\quad{}\quad{}\quad{}\quad{}\quad{}\quad{}\quad{}\quad{}\quad{}\quad{}\textbf{(b)}

\bigskip{}
\bigskip{}

\includegraphics[width=0.45\columnwidth]{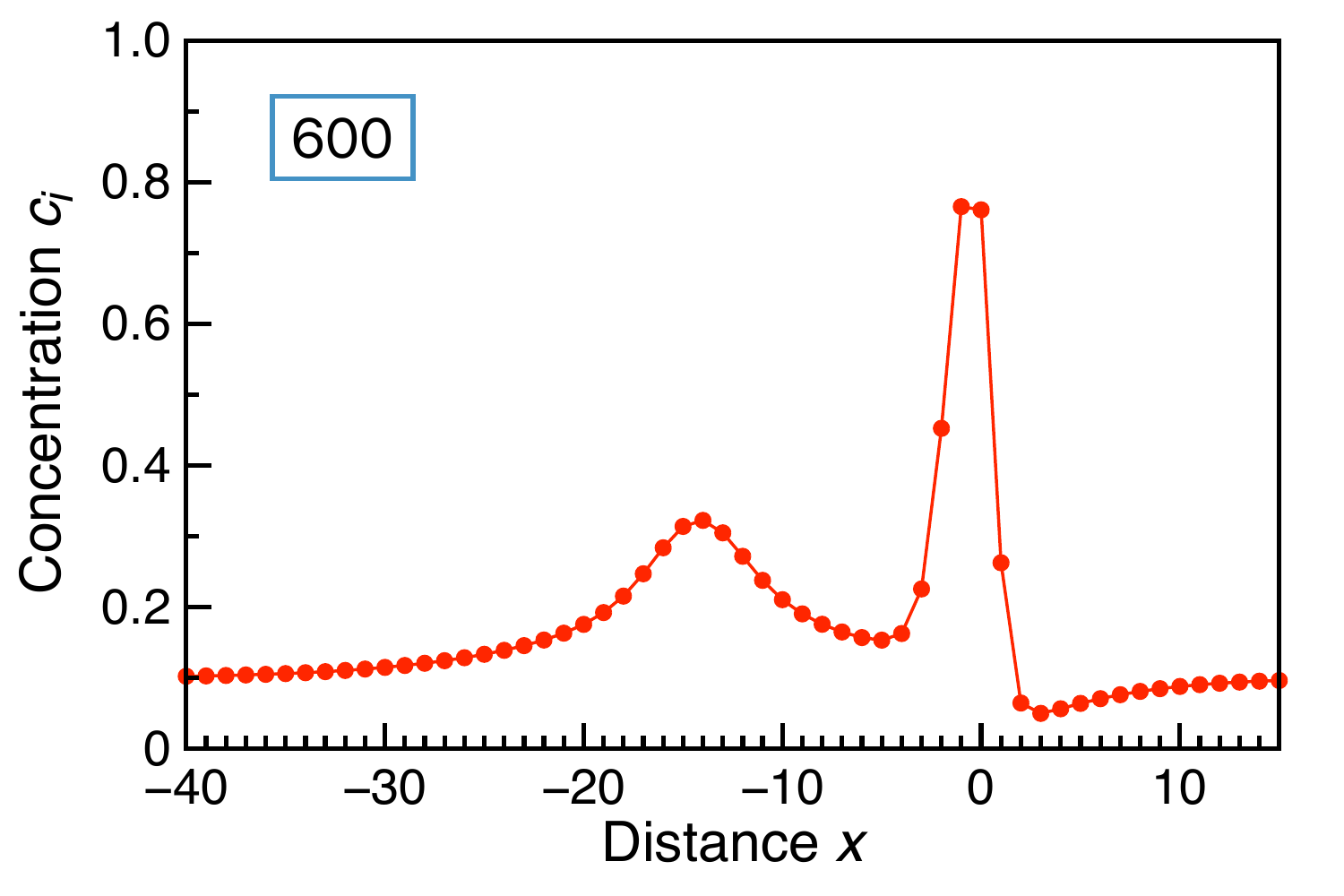}\quad{}\includegraphics[width=0.45\columnwidth]{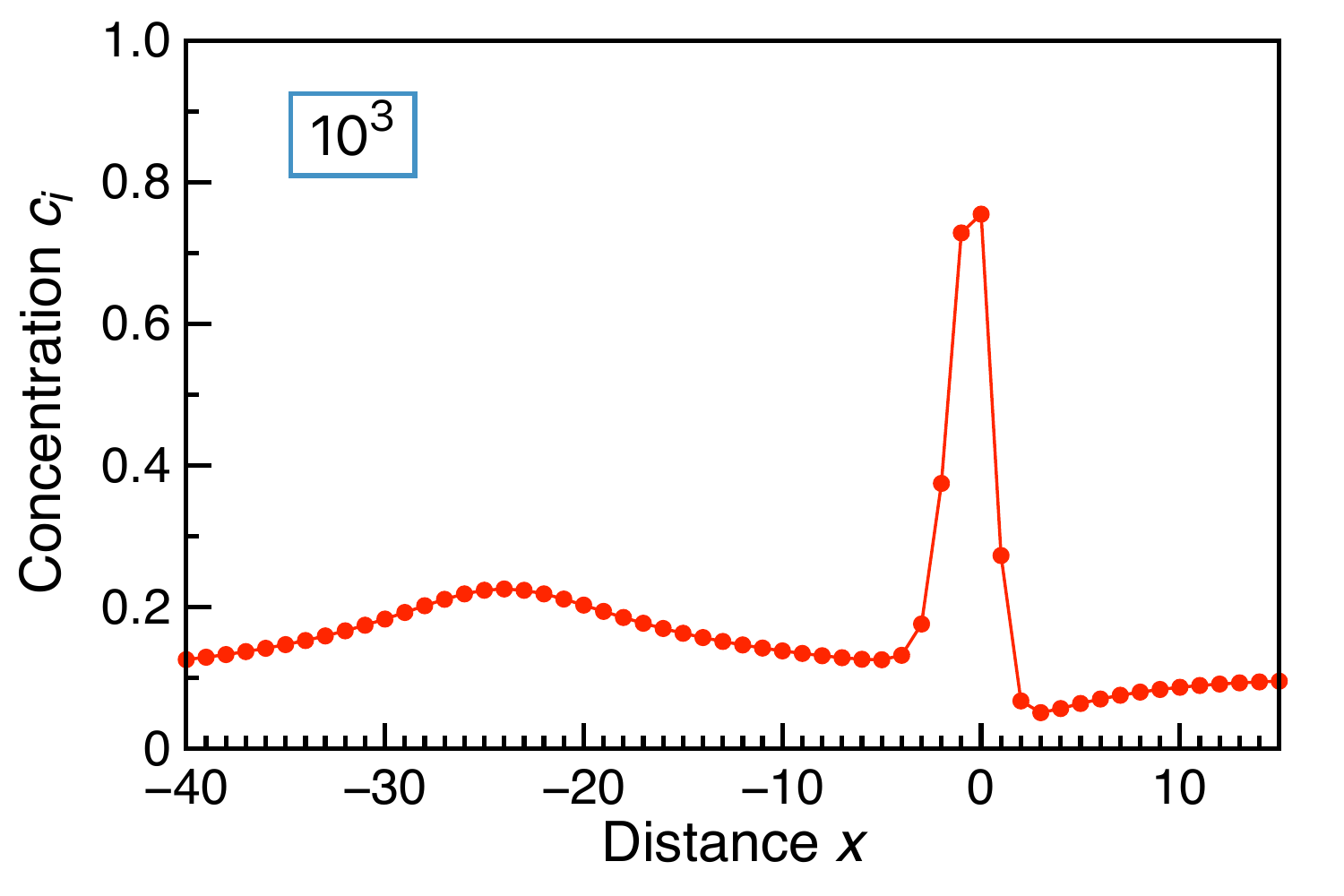}

\textbf{(c)}\quad{}\quad{}\quad{}\quad{}\quad{}\quad{}\quad{}\quad{}\quad{}\quad{}\quad{}\quad{}\quad{}\quad{}\quad{}\quad{}\quad{}\quad{}\textbf{(d)}

\bigskip{}
\bigskip{}

\includegraphics[width=0.45\columnwidth]{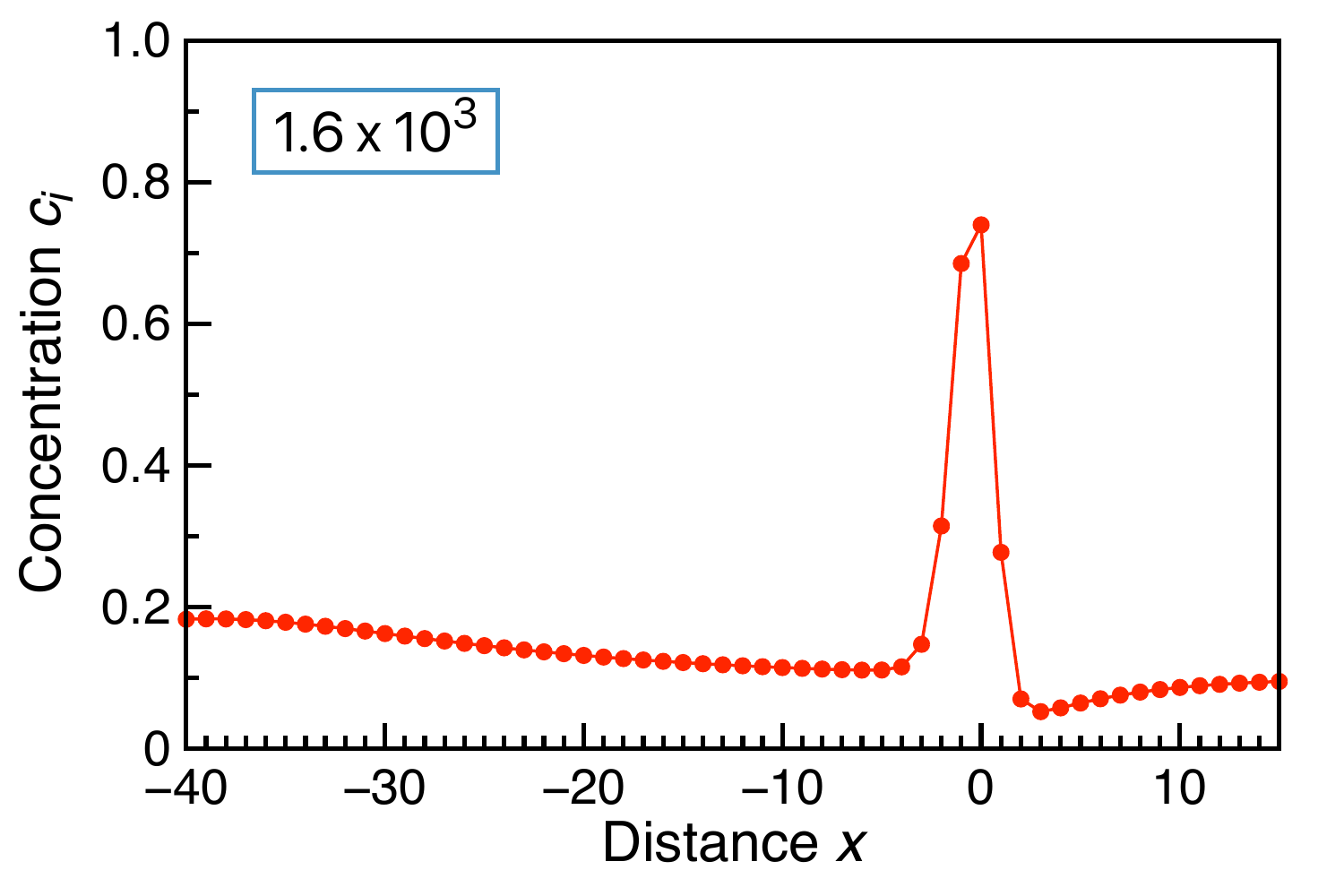}\quad{}\includegraphics[width=0.45\columnwidth]{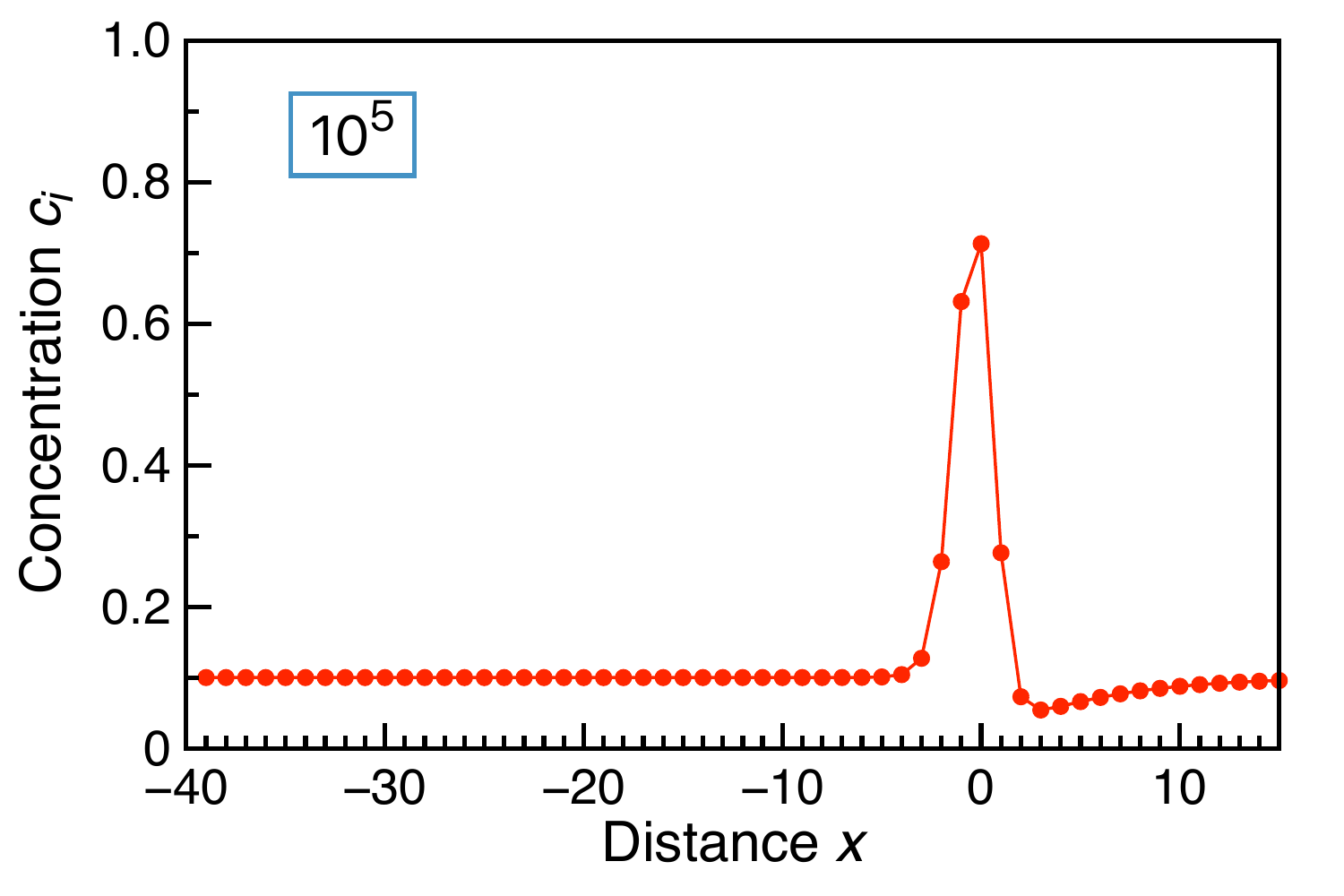}

\textbf{(e)}\quad{}\quad{}\quad{}\quad{}\quad{}\quad{}\quad{}\quad{}\quad{}\quad{}\quad{}\quad{}\quad{}\quad{}\quad{}\quad{}\quad{}\quad{}\textbf{(f)}

\caption{Time evolution of the segregation atmosphere for the alloy composition
$c=0.1$ and temperature $T=0.8$. After the grain boundary was equilibrated
at $V=0$, the speed $V=0.1$ was instantaneously applied and remained
constant. The solute composition is plotted as a function of distance
$x$ from the grain boundary position at the moments of time indicated
in the key. The reference frame is attached to the moving grain boundary.\label{fig:Time-evolution-1}}

\end{figure}

Simulations with instantaneous application of a nonzero speed to an
equilibrium grain boundary provide another interpretation of the right-
and left-hand branches of the solute drag curve (Fig.~\ref{fig:f-V_1}).
If the grain boundary is first equilibrated at $V=0$ and then a speed
$V>0$ lying on the left-hand branch is applied, the segregation profile
evolves into the one that would be obtained by slowly increasing the
speed from zero to $V$. In other words, the segregation atmosphere
catches up with the sudden onset of the grain boundary motion and
starts moving together with the boundary. By contrast, application
of a speed lying on the right-hand branch causes separation of the
boundary from the atmosphere. As soon as the boundary breaks away
from the initial atmosphere, it start forming a new one, which eventually
develops into a steady state segregation profile corresponding to
the applied speed. The new segregation peak is always lower and more
narrow than the old. The process is illustrated in Fig.~\ref{fig:Time-evolution-1}
showing the evolution of the composition profile after a speed $V=0.1$
lying on the right-hand branch was applied. The initial atmosphere
left behind the boundary creates a compositional peak inside the advancing
grain. A grain boundary atmosphere without the grain boundary is reminiscent
of the grin of the Cheshire Cat, which remained for some time after
the rest of the Cat had vanished \citep{Carroll:1865aa}. This phenomenon
creates so-called ``ghost'' grain boundaries in materials that can
be observed experimentally. The ``ghost'' atmosphere spreads out by
solute diffusion and eventually disappears, as illustrated in Fig.~\ref{fig:Time-evolution-1}(f).

The same breakaway phenomenon was observed when the grain boundary
was moved slowly with a speed lying on the left-hand branch and then
the speed was suddenly increased to a value lying on the right-hand
branch. In some of the simulations, the two speeds were chosen so
that they lied on different branches but corresponded to the same
drag force. For example, a boundary moving with the speeds 0.009 and
0.1 experiences the same drag force of about 0.194. Thus, the switch
from the left-hand branch to the right-hand branch is accompanied
by a drastic increase in the grain boundary mobility.\footnote{Grain boundary mobility is often defined as the proportionality factor
between the driving force and the speed, assuming the relation is
linear. In this paper we use this term in a more generic sense. We
say that the mobility increases if the boundary moves faster under
the same driving force, or requires a smaller force for the motion
with the same speed.} The boundary sheds a ``heavy'' atmosphere and forms a ``lighter''
one (smaller segregation) that allows it to move an order of magnitude
faster under the same driving force.

\subsection{Phase transformations in a moving grain boundary}

So far we have only considered the part of the solute drag curve on
which the grain boundary properties are smooth functions of the speed
as the latter increases starting from zero (Figs.~\ref{fig:f-V_1}
and \ref{fig:Properties_A}). Since the grain boundary state at $V=0$
corresponds to phase $\beta$, we can consider the moving boundary
as being a dynamic version of the phase $\beta$. This interpretation
is consistent with the definition of a phase as a continuum of states
whose properties are described by a smooth function of control variables.
Furthermore, the dynamic properties of the $\beta$ phase are different
on the right-hand and left-hand branches of the solute drag curve.
To reflect this difference, we can refer to the branches as the ``fast
$\beta$'' and ``slow $\beta$'' phases, respectively.

As the boundary speed increases, a point is reached at which the dynamic
phase $\beta$ abruptly switches to a new state of motion, which is
characterized by significantly different properties. The amount of
segregation decreases, the grain boundary free energy increases, and
the drag force drops to a much smaller value. As the speed increases
further, all grain boundary properties remain smooth functions of
the speed, suggesting that the boundary has transformed to a new phase.
This phase can be identified as the dynamic version of the $\alpha$
phase since its $\Gamma$ and $\gamma$ are close to those for the
equilibrium $\alpha$ phase. In fact, since the drag force acting
on this phase decreases with the speed, this is the fast version of
the $\alpha$ phase.

Simulations were also performed by gradually decreasing the speed
of the $\alpha$ phase while keeping it in the steady state mode.
It was found that, at a certain speed, the boundary spontaneously
switches back to the $\beta$ phase. This $\alpha\rightarrow\beta$
transition occurs at a smaller speed than the $\beta\rightarrow\alpha$
transition, creating a hysteresis effect. The hysteretic behavior
of the grain boundary properties is illustrated by the red and blue
curves in Figs.~\ref{fig:f-V_1} and \ref{fig:Properties_A}. A similar
behavior was observed by Ma et al.~\citep{Wang03} in their phase-field
model.

\begin{figure}
\noindent \begin{centering}
\includegraphics[width=0.515\textwidth]{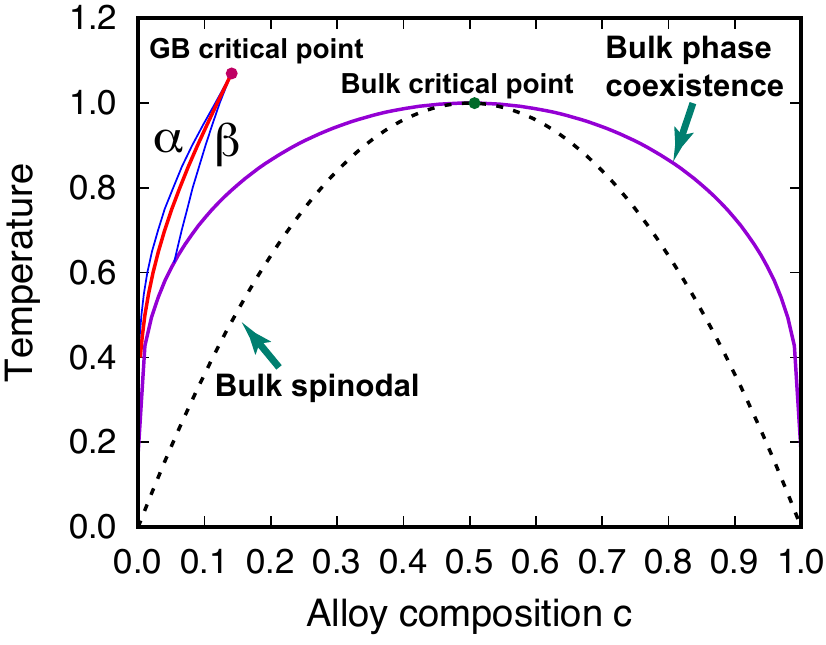}\quad{}\includegraphics[width=0.415\textwidth]{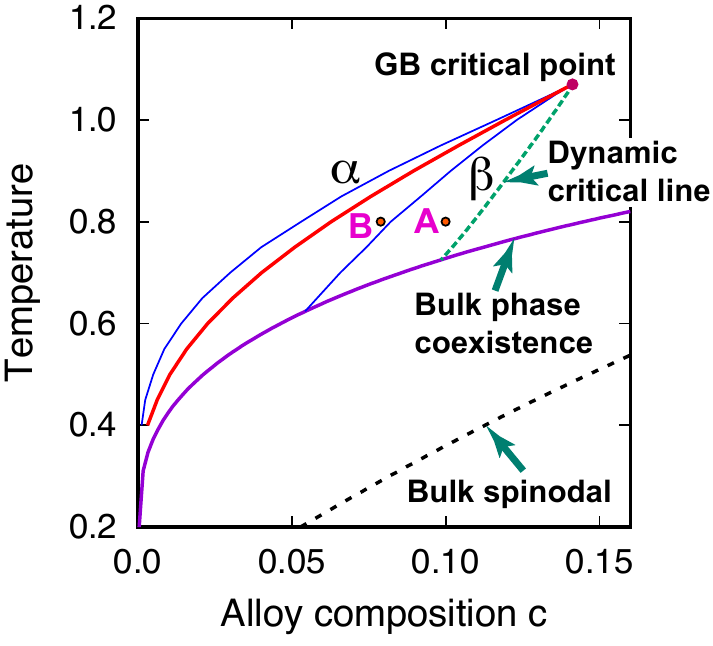}
\par\end{centering}
\noindent \begin{centering}
\textbf{(a)}\quad{}\quad{}\quad{}\quad{}\quad{}\quad{}\quad{}\quad{}\quad{}\quad{}\quad{}\quad{}\quad{}\textbf{(b)}
\par\end{centering}
\caption{Grain boundary phase transformation lines superimposed on the bulk
phase diagram. The grain boundary phase coexistence line (red) and
spinodal lines (blue) merge at a critical point. (a) Full diagram.
(b) Zoom into the important area showing the alloys A and B discussed
in the text. The dynamic critical line of the grain boundary phases
is indicated.\label{fig:Phase_diag}}
\end{figure}

These results demonstrate an interesting phenomenon of dynamic transformations
between grain boundary phases accompanied by a dynamic hysteresis.
This behavior looks similar to equilibrium phase transformations,
seen for example in Figs.~\ref{fig:Equi_1}-\ref{fig:Phase_diag}.
The difference is that the role of the control parameter is played
by the grain boundary speed, instead of the thermodynamic variables
such as chemical composition or temperature. Another remarkable observation
is that the chosen alloy lies outside the grain boundary spinodal
line on the phase diagram (Fig.~\ref{fig:Phase_diag}). As such,
the $\alpha$ phase is absolutely unstable at $V=0$. If initially
created in the grain boundary, it spontaneously transforms to the
$\beta$ phase without any thermodynamic barrier. However, grain boundary
motion brings this phase to existence. In fact, at large enough speeds
this phase is the only possible state of  steady state motion of the
boundary.

\begin{figure}
\begin{centering}
\textbf{(a)}\enskip{}\includegraphics[width=0.5\textwidth]{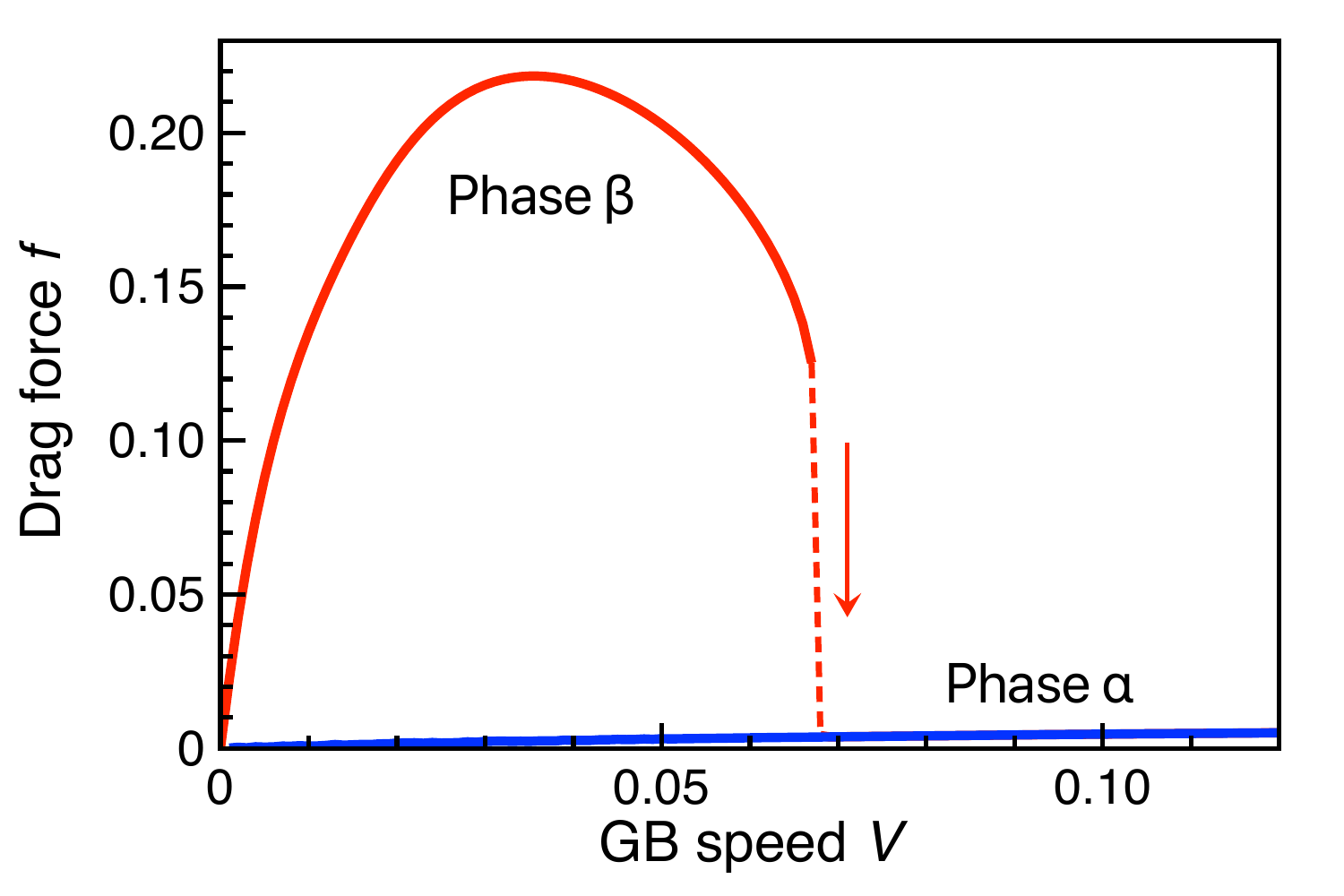}
\par\end{centering}
\bigskip{}

\begin{centering}
\textbf{(b)}\enskip{}\includegraphics[width=0.52\textwidth]{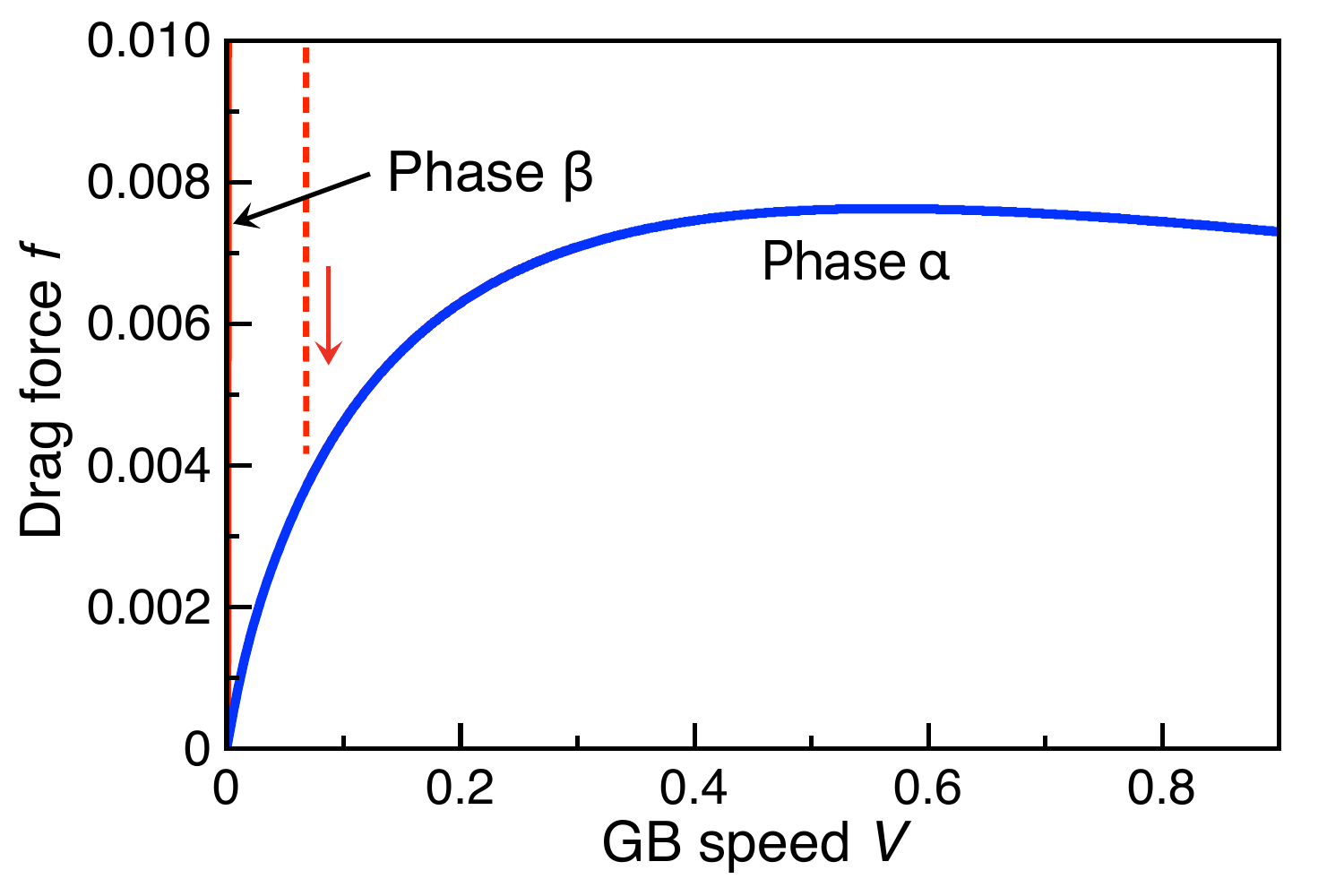}
\par\end{centering}
\caption{(a) Solute drag force $f$ as a function of grain boundary speed $V$
for the alloy composition $c=0.08$ and temperature $T=0.8$ (alloy
B in Fig.~\ref{fig:Phase_diag}(b)). The red and blue curves were
obtained by, respectively, increasing and decreasing the speed. The
arrow indicates the transformation from phase $\beta$ to phase $\alpha$.
(b) Zoom into the $\alpha$ phase curve.\label{fig:f-V_2}}

\end{figure}

A somewhat different scenario of the phase transformations is observed
in the alloy with $c=0.1$ and $T=0.8$, corresponding to point B
on the phase diagram (Fig.~\ref{fig:Phase_diag}(b)). As above, phase
$\beta$ is the most stable state of the boundary, but phase $\alpha$
is now metastable. We can equilibrate the boundary in either phase
and use this as the initial state for the boundary motion. We thus
obtain two force-speed functions, one for each phase (Fig.~\ref{fig:f-V_2}).
Their plots have the familiar shape with a maximum separating the
slow and fast versions of each phase. A striking feature is the disparity
in the scales of the two curves: the peak of the $\alpha$ phase is
so much lower than the peak of the $\beta$ phase that they can hardly
be shown on the same plot. The difference between the mobilities of
the two phases becomes even greater as we further decrease the solute
concentration in the grains and/or the temperature.

If the initial state is phase $\alpha$, the solute drag curve extends
all the way to infinity along the speed axis. Reversal of the speed
retraces the same curve. This behavior can be described as ``dynamically
reversible''. By contrast, as the speed increases, the $\beta$ phase
eventually loses its dynamic stability and transforms to the $\alpha$
phase (Fig.~\ref{fig:f-V_2}). In other words, despite being more
stable thermodynamically, the $\beta$ phase becomes less stable than
the $\alpha$ phase when the boundary moves. Once the $\beta\rightarrow\alpha$
transition has occurred, the boundary does not return to the $\beta$
phase at any speed. As in the case of alloy A, the $\alpha$ phase
is the only possible state of the grain boundary motion in the high-speed
limit. Figure \ref{fig:Properties_B-1} demonstrates the discontinuities
in the grain boundary properties accompanying the $\beta\rightarrow\alpha$
transformation.

\subsection{The dynamic critical line on the phase diagram}

\begin{figure}
\noindent \begin{centering}
\includegraphics[width=0.6\textwidth]{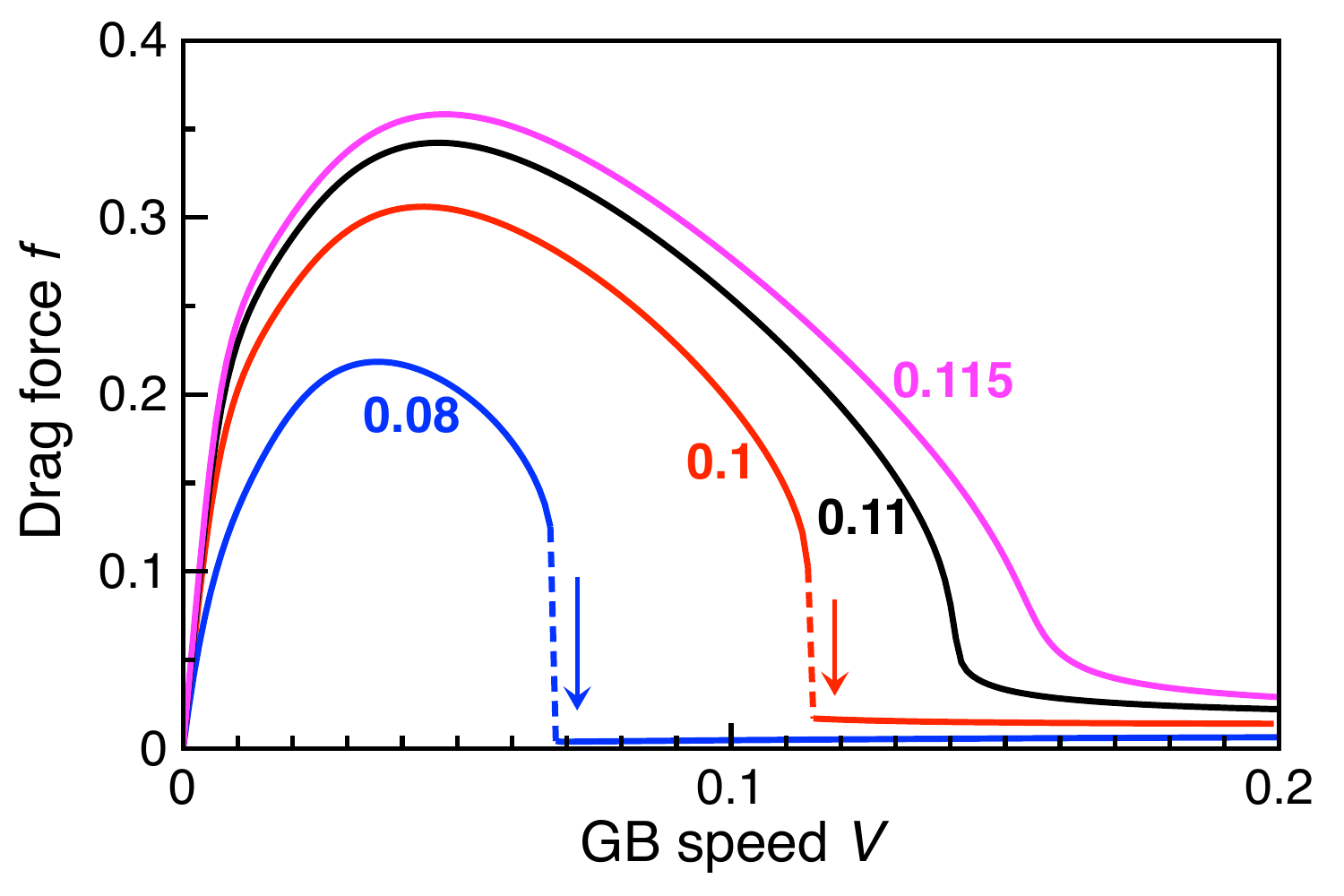}
\par\end{centering}
\caption{Solute drag force $f$ as a function of grain boundary speed $V$
for a set of alloys at the temperature $T=0.8$. The alloys compositions
are indicated by the labels. The initial grain boundary state is the
equilibrium phase $\beta$. The arrows mark discontinuities caused
by the dynamic phase transformation $\beta\rightarrow\alpha$. \label{fig:Critical_behavior}}

\end{figure}

The two alloys A and B discussed in the previous sections (Fig.~\ref{fig:Phase_diag}(b))
only serve to illustrate some typical dynamic regimes of the solute
drag. Simulations conducted for other alloy compositions and temperatures
revealed additional interesting features, one of which being the existence
of a dynamic critical line. Figure~\ref{fig:Critical_behavior} shows
a set of solute drag curves obtained by increasing the speed starting
from the equilibrium $\beta$ phase. The plot illustrates the evolution
of the curves with increasing alloy composition at a fixed temperature
($T=0.8$). Note that the jump in the solute drag force decreases
as the solute concentration in the grains increases, and eventually
shrinks to zero at $c=0.11$. At higher concentrations, the $\beta\rightarrow\alpha$
transformation is continuous.

This behavior indicates that $c=0.11$ is a dynamic critical point
of the $\beta\rightarrow\alpha$ transformation at this temperature.
The loci of such critical points form a dynamic critical line in the
space of variables ($T$, $c$, $V$). This line starts at the equilibrium
critical point ($1.072$, $0.1418$, $0$) and continues towards lower
temperatures and larger concentrations and speeds. A projection of
this line on the $c$-$T$ phase diagram is shown in Fig.~\ref{fig:Phase_diag}(b).
In all alloys on the right of this line, the dynamic $\beta\rightarrow\alpha$
transformation occurs continuously. We emphasize again that the dynamic
$\alpha$ phase resulting from this transformation, whether it occurs
continuously or discontinuously, is absolutely unstable $V=0$.

\section{Discussion and conclusions}

The discrete model discussed in this paper relies on many simplifying
assumptions and approximations. It is assumed that the grain boundary
does not have any particular structure different from the structure
of the grains. In the future, some elements of a grain boundary structure
could be introduced by making the coordination numbers $z$ and $z^{\prime}$
in the boundary region different from those in the grains. In this
paper, however, the grain boundary-solute interactions are treated
as purely chemical. In a single-component state (pure A or pure B),
the grain boundary has an excess energy due to the modified strength
of the chemical bonds within the grain boundary region. It should
be noted, however, that this modified bond energetics is not coupled
to the boundary motion. It is assumed that the bonds can quickly change
their strength during the boundary motion without any dissipation.
In other words, the Gaussian describing the modified bond energies
moves without any resistance. As a consequence, a single-component
grain boundary has an infinitely high mobility. The resistance to
its motion only comes from interactions with the solute atoms, which
move with a finite rate controlled by diffusion. While an intrinsic
resistance term can be easily added to the drag force, its speed dependence
and all other dynamic characteristics cannot be predicted by the present
model.

Another limitation of the model is the simplified thermodynamic treatment.
The atoms are assumed to interact by nearest-neighbor bonds on a rigid
lattice, neglecting long-range forces, many-body effects, structural
relaxation and elastic interactions. The regular solution model for
the free energy is another strong approximation. Further, the grain
boundary is assumed to remain perfectly planar and move with a constant
speed. Thus, many interesting phenomena associated with morphological
evolution of moving grain boundaries are left outside the model. Similar
to the classical models by Cahn \citep{Cahn1962} and Lücke et al.~\citep{Lucke-Stuwe-1963,Lucke:1971aa},
the planar geometry of the present model makes it essentially one-dimensional.
The equations describing the grain boundary dynamics are purely deterministic,
which precludes direct observation of the phase nucleation process
and thus two-phase states of the grain boundary. To observe the nucleation,
the model would need to be extended to three dimensions and include
thermal fluctuations.

At the same time, the model offers a number of significant advantages
over alternative approaches. By treating the solute diffusion process
phenomenologically, the model overcomes the time-scale limitation
of MD and affords solute drag simulations over a wide range of time,
distances and speeds. This enables us to implement different dynamic
regimes of the solute drag process. The computational efficiency of
the models permits broad exploration of the parameter space and observation
of both the transient regime and the steady state motion for each
set of parameters. The grain boundary thermodynamics and kinetics
are treated within the same unified framework, making the model internally
consistent. The model provides easy access to the grain boundary free
energy and the solute drag force, whose calculation from MD simulations
would be extremely challenging. By contrast to the diffuse-interface
treatments \citep{Wang03,Li:2009aa,Shahandeh:2012aa,Gronhagen:2007aa,Abdeljawad:2017aa},
this model preserves the discrete character of crystalline systems
and does not rely on the gradient approximation. Despite its approximate
character, the model captures the essential physics of the solute
drag process.

\begin{figure}
\noindent \begin{centering}
\includegraphics[width=0.55\textwidth]{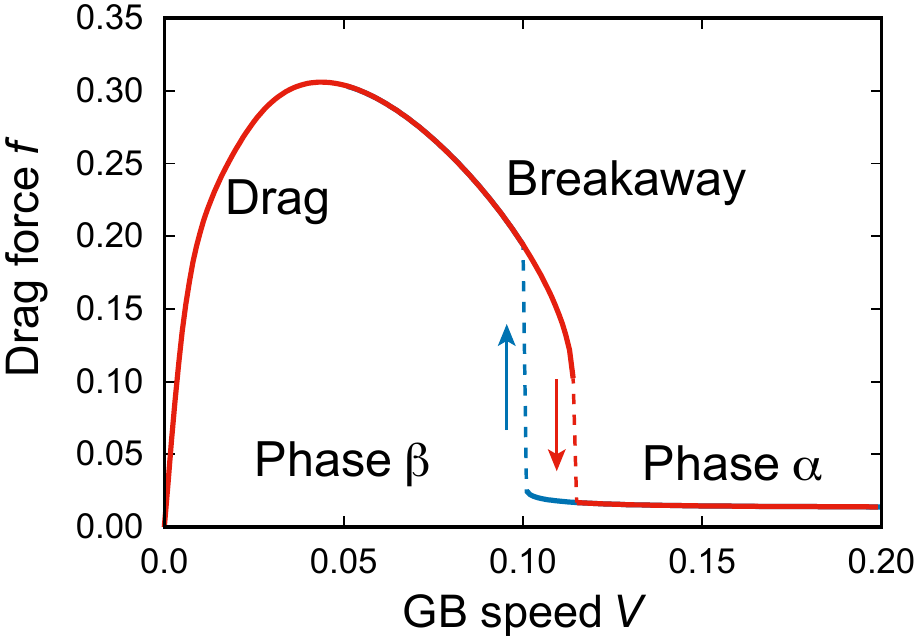}
\par\end{centering}
\caption{Solute drag force $f$ as a function of grain boundary speed $V$
for the alloy composition $c=0.1$ and temperature $T=0.8$ (alloy
A in Fig.~\ref{fig:Phase_diag}(b)). The red and blue curves were
obtained by gradually increasing and decreasing the speed, respectively.
The transformation between the $\alpha$ and $\beta$ phases is accompanied
by a hysteresis loop shown by the dashed lines.\label{fig:f-V_1}}

\end{figure}

As mentioned earlier, the high-speed branch of the solute drag curve
(Fig.~\ref{fig:f-V_1}), where the drag force decreases with the
speed, is often dismissed as being associated with morphologically
unstable grain boundary motion. This instability was predicted by
Cahn \citep{Cahn1962} based on qualitative arguments and was later
analyzed more rigorously within more general models that permit variations
in the grain boundary shape and lateral diffusion of the solute \citep{Roy:1975aa,Korzhenevskii:2002aa,Korzhenevskii:2006aa}.
In the high-speed limit, this instability becomes similar to the Mullins-Sekerka
instability for a planar solid-liquid interface during solidification
\citep{Mullins:1964aa}. In this paper we treat this mode of grain
boundary motion on equal footing with the stable (low speed) branch
of the solute drag curve. The stability analyses mentioned above are
based on many simplifying assumptions, ranging from the dilute-solution
approximation to the assumption of isotropic interface tension. We
cannot exclude that, under certain conditions, the moving boundary
can preserve its nearly planar shape even if the drag force decreases
with the speed. For example, if the interface tension is strongly
anisotropic and its plane orientation corresponds to a cusp in the
angular dependence of tension, then Herring's torque term \citep{Herring:1951aa}
will resist the development of protrusions and other changes in the
morphology. Under such conditions, the high-speed branch of the solute
drag curve can be quite relevant to the grain boundary motion in real
materials.

One of the interesting findings of this work is that the grain boundary
motion preserves the existence of and the transformations between
the grain boundary phases. In fact, the grain boundary motion can
even stabilize absolutely unstable phases, i.\,e., phases that must
spontaneously transforms to a more stable phase in a stationary grain
boundary. The dynamic phase transformations found in this paper exhibit
many features of equilibrium phase transformations, such as the dynamic
hysteresis and a dynamic critical line. The grain boundary phase diagram
can be extended to moving grain boundaries, with the steady state
speed $V$ playing the role of an additional control parameter. For
a binary system, the grain boundary spinodal lines become spinodal
surfaces in the $(c,T,V)$ space of variables. These surfaces intersect
along the dynamic critical line, which emanates from the equilibrium
critical point.

There are two fundamental issues that could not be resolved in this
paper and are left for future work. One is related to the thermodynamic
meaning of the excess properties for moving grain boundaries. For
an equilibrium grain boundary, its excess free energy $\gamma$ defined
by Eq.(\ref{eq:gamma}) follows the Gibbs adsorption equation \citep{Willard_Gibbs}
\begin{equation}
d\gamma=-SdT+vdp-\Gamma d\mu,\label{eq:adsorption}
\end{equation}
where the segregation $\Gamma$ is given by Eq.(\ref{eq:segreg})
and $S$ and $v$ are the excess entropy and volume, respectively.
$S$ is defined in a similar manner as $\gamma$ and can be readily
computed in this model. $v$ is identically zero in this model due
to the rigid lattice approximation. As a check of our methodology,
the relation
\[
\Gamma=-\left(\dfrac{\partial\gamma}{\partial\mu}\right)_{T,V=0}
\]
was verified by numerical calculations for a wide range of the model
parameters.

\begin{figure}
\noindent \begin{centering}
\includegraphics[width=0.6\textwidth]{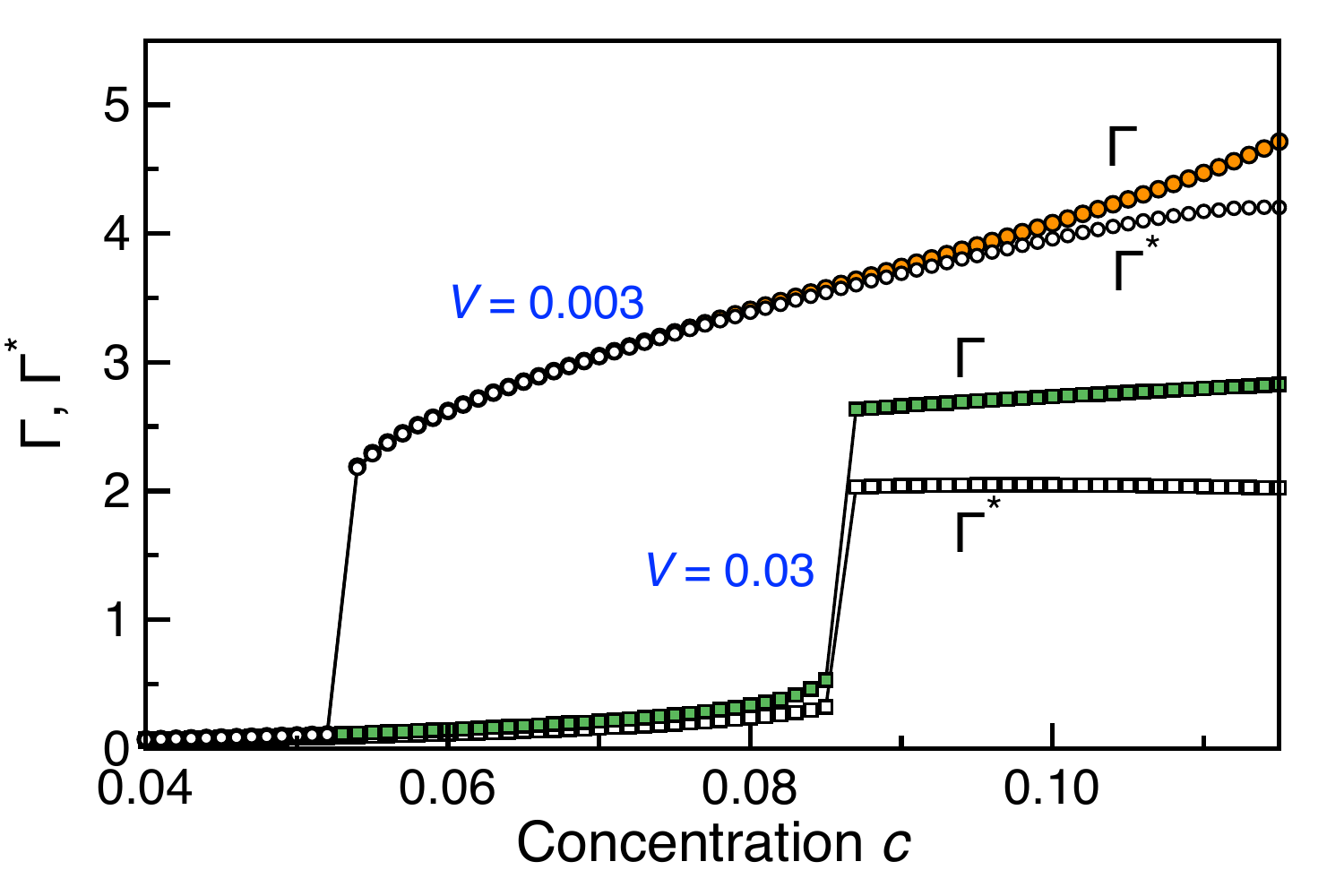}
\par\end{centering}
\caption{Grain boundary segregation $\Gamma$ (filled symbols) and the derivative
$\Gamma^{*}=-(\partial\gamma/\partial\mu)_{T,V}$ (open symbols) as
functions of the alloy composition at the temperature $T=0.8$ for
two values of the grain boundary speed $V$. The discontinuities are
caused by the grain boundary phase transformations. The alloy composition
was decreased by small increments while keeping the boundary motion
in a steady state mode.\label{fig:Compare_Gammas}}

\end{figure}

Now consider a grain boundary moving in a steady state mode. An observer
moving with the grain boundary speed $V$ will see a stationary composition
profile and will be able to compute all excess quantities using the
same equations as for an equilibrium boundary. These excess quantities
will be functions of $T$, $\mu$ and $V$. The question then arises
as to whether the quantities thus obtained are related to each other
through a generalized form of Eq.(\ref{eq:adsorption}). Since $\gamma=\gamma(T,\mu,V)$,
we can write down its differential
\begin{equation}
d\gamma=-S^{*}dT-\Gamma^{*}d\mu+\eta dV,\label{eq:adsorption-1}
\end{equation}
where $-S^{*}$, $-\Gamma^{*}$ and $\eta$ denote the respective
partial derivatives of $\gamma$. One would hope that $S^{*}$ and
$\Gamma^{*}$ coincide with the dynamic values of $S$ and $\Gamma$
seen by the moving observer. This would make Eq.(\ref{eq:adsorption-1})
a generalized form of the adsorption equation for moving grain boundaries.
Unfortunately, numerical tests show that this is not the case. For
example, Fig.~\ref{fig:Compare_Gammas} compares $\Gamma$ and $\Gamma^{*}=-(\partial\gamma/\partial\mu)_{T,V}$
as functions of the alloy composition at fixed values of $T$ and
$V$. While the two functions approach each other in the limit of
small speeds, they are generally different. Thus, the differential
coefficients in Eq.(\ref{eq:adsorption-1}) are generally not the
excess quantities predicted by the standard Gibbsian equations such
as Eq.(\ref{eq:segreg}).

\begin{figure}
\noindent \begin{centering}
\includegraphics[width=0.6\textwidth]{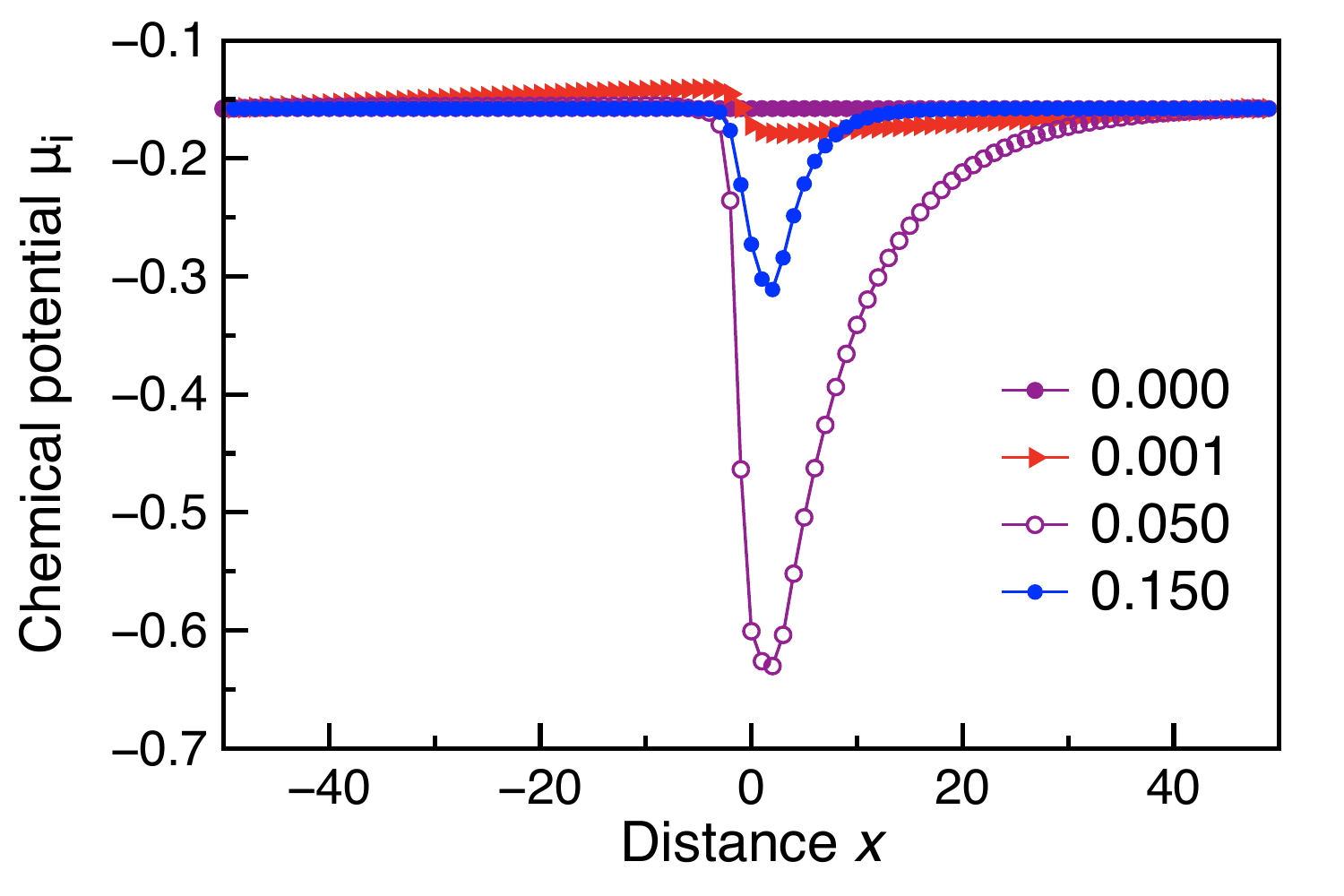}
\par\end{centering}
\caption{The chemical potential as a function of distance for a grain boundary
moving with the speeds indicated in the key at the temperature $T=0.8$
and alloy composition $c=0.1$. \label{fig:Chemical-potential}}
\end{figure}

It follows that the excess quantities, including the excess free energy,
must be redefined for a moving grain boundary. The apparent similarity
between the equilibrium grain boundary and the stationary grain boundary
seen by the moving observer is deceiving. Some of the grain boundary
properties are fundamentally different. As one example, while the
chemical potential is uniform across an equilibrium grain boundary,
it is highly non-uniform inside the segregation atmosphere of a moving
boundary, as illustrated in Fig.~\ref{fig:Chemical-potential}. (This
could be a clue to redefining the excess free energy using the coordinate-dependent
chemical potential.) This and similar examples suggest that a future
theory of moving grain boundaries can be more complex than the equilibrium
interface thermodynamics. Before such a theory exists, the dynamic
free energy $\gamma$ defined by Eq.(\ref{eq:gamma}) can only be
utilized as a useful indicator of phase transformations based on its
discontinuities, but cannot inserted in any adsorption equation.

\begin{figure}
\begin{centering}
\includegraphics[width=0.55\textwidth]{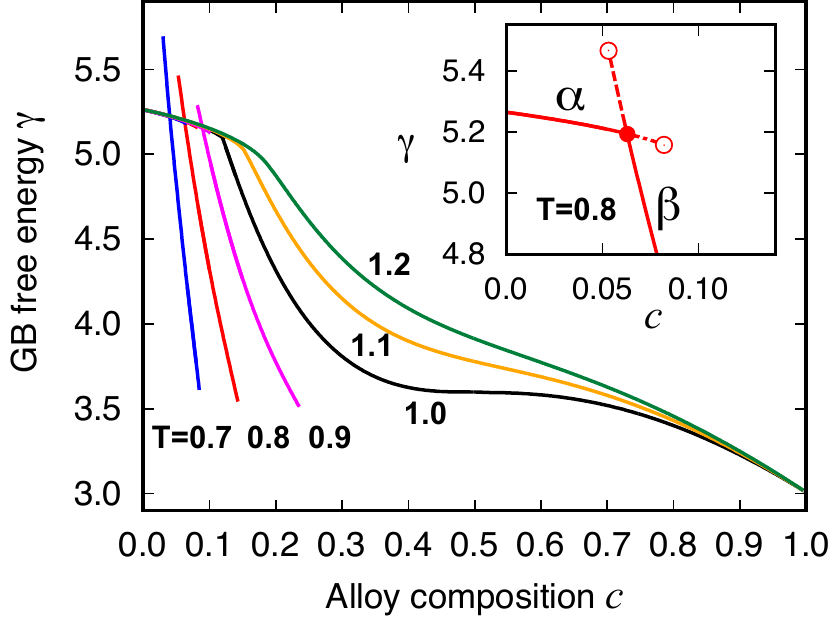}
\par\end{centering}
\caption{Isotherms of the grain boundary free energy $\gamma$ at selected
temperatures $T$. The inset is a zoom into the phase transformation
region at $T=0.8$. The dashed lines represents metastable states,
the filled circle the grain boundary phase equilibrium, and the open
circles the grain boundary spinodal points.\label{fig:gamma_plot}}

\end{figure}

\begin{figure}
\noindent \begin{centering}
\textbf{(a)}\enskip{}\includegraphics[width=0.42\textwidth]{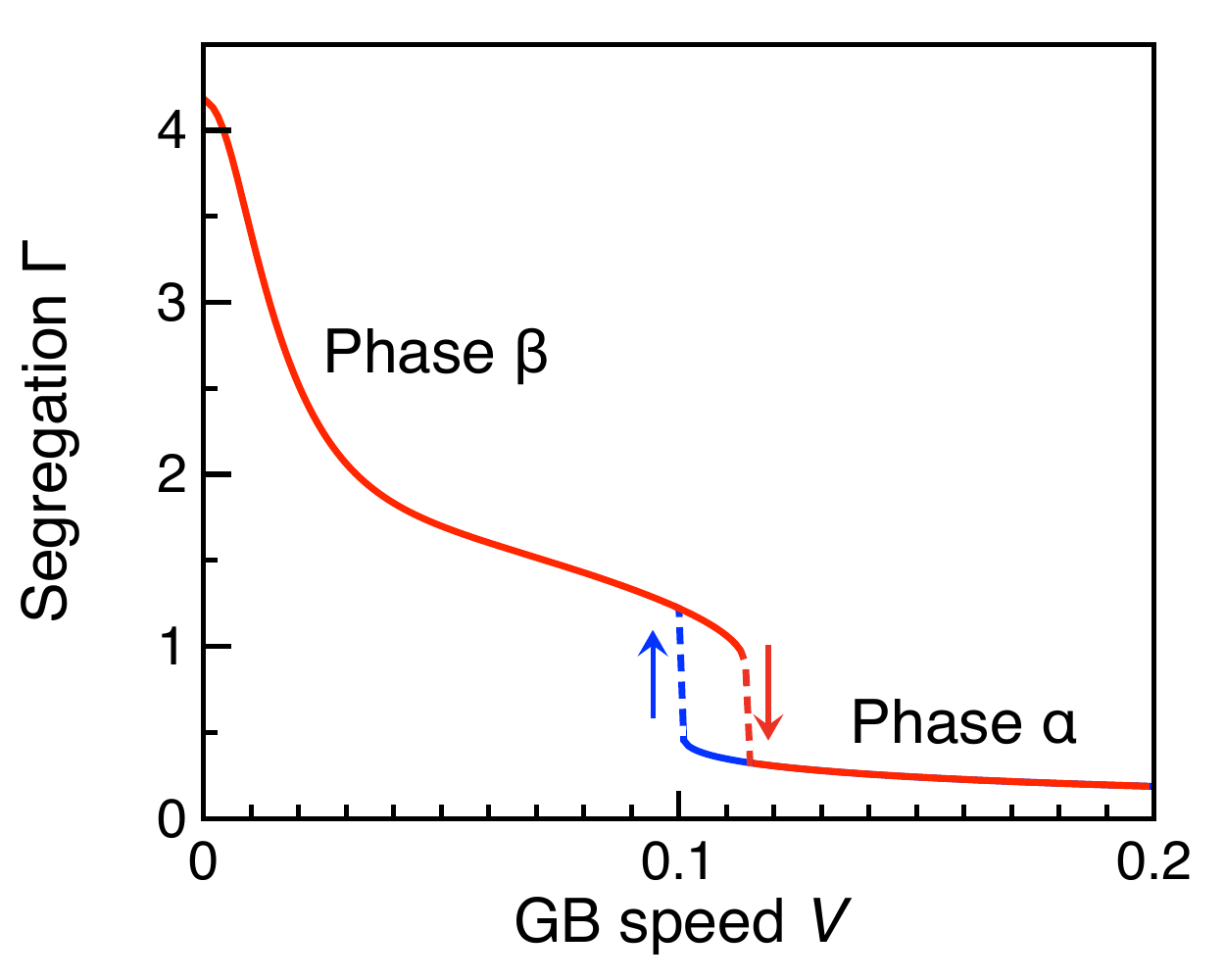}
\par\end{centering}
\noindent \begin{centering}
\bigskip{}
\par\end{centering}
\noindent \begin{centering}
\textbf{(b)}\enskip{}\includegraphics[width=0.42\textwidth]{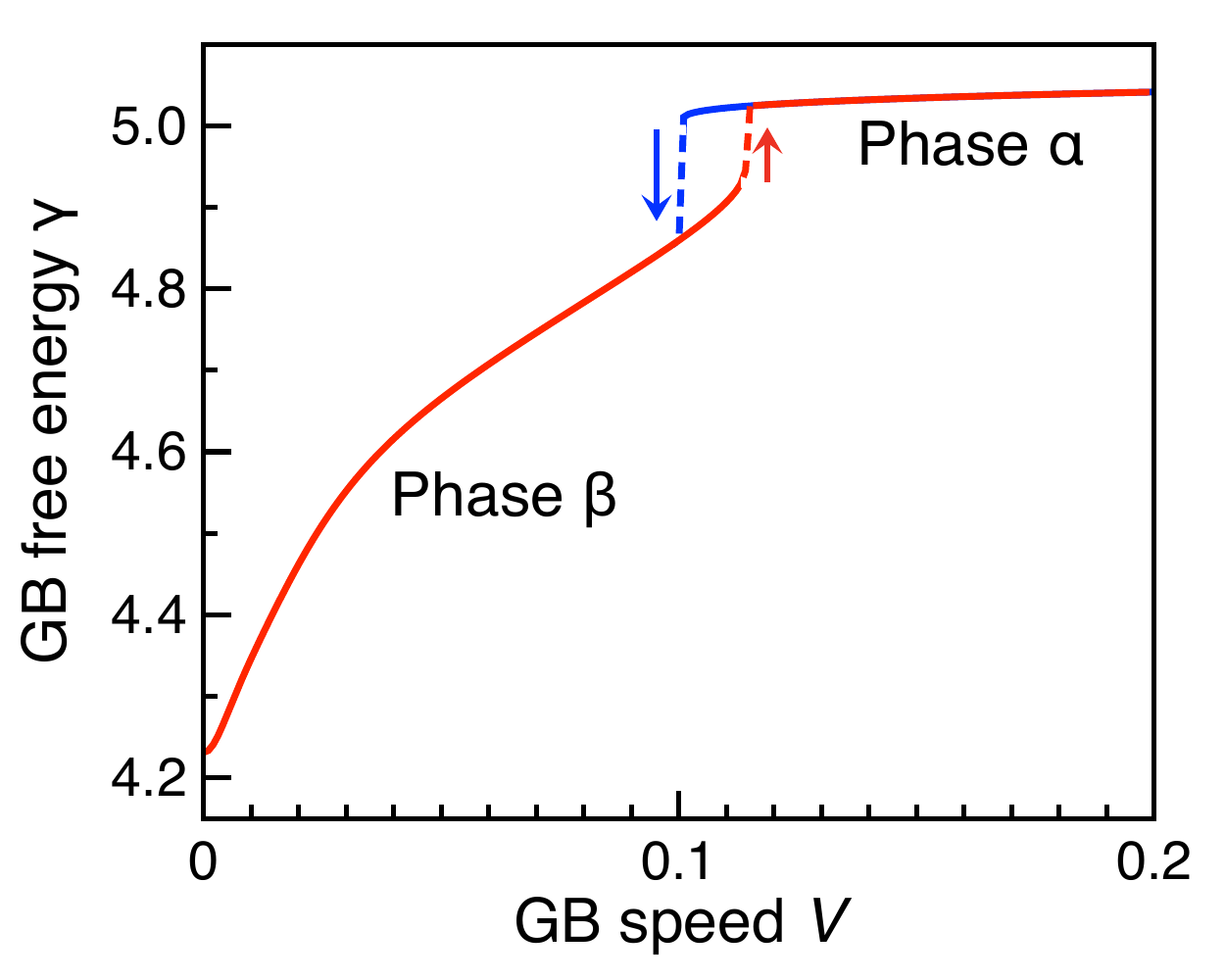}
\par\end{centering}
\noindent \begin{centering}
\bigskip{}
\par\end{centering}
\noindent \begin{centering}
\textbf{(c)}\enskip{}\includegraphics[width=0.42\textwidth]{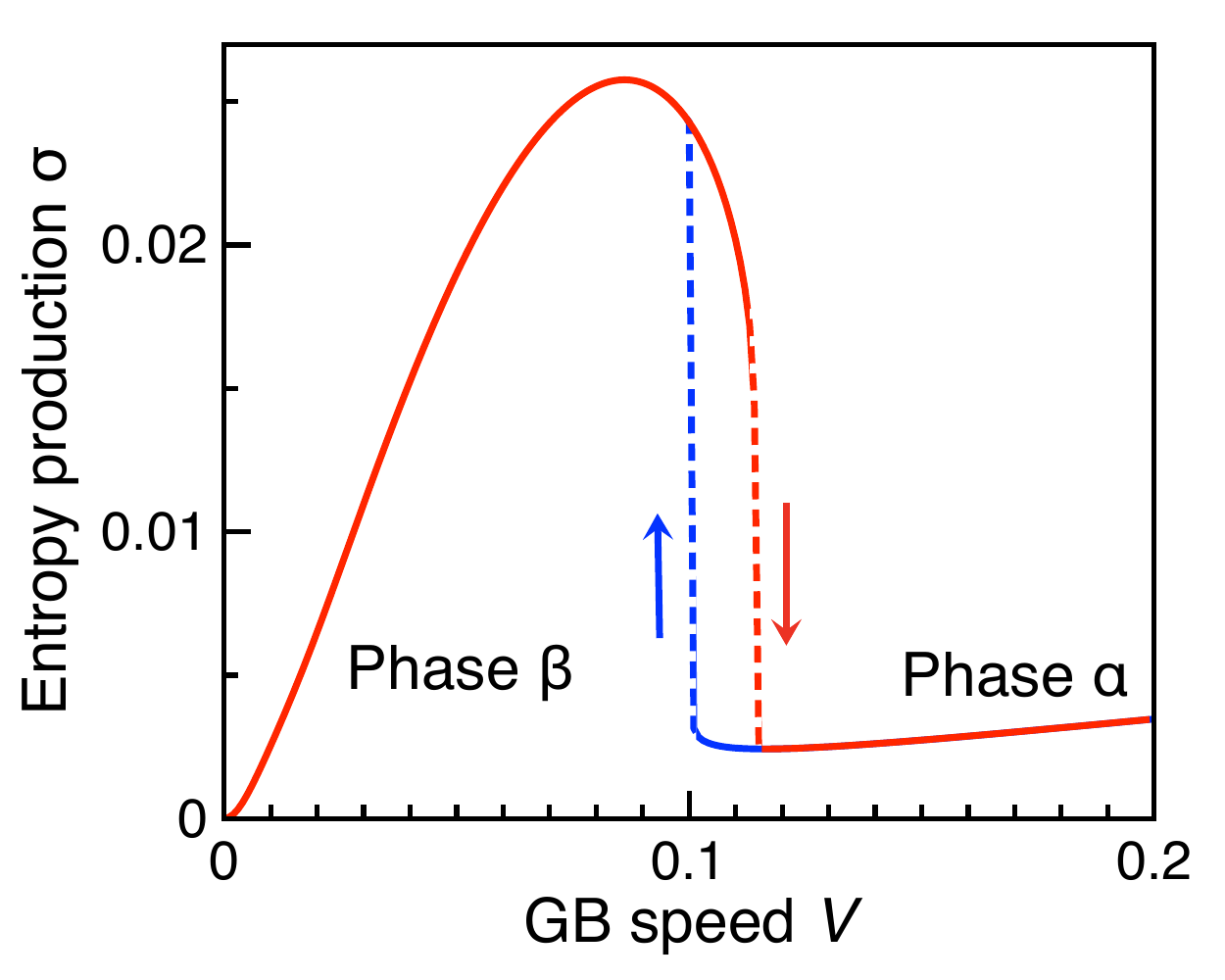}
\par\end{centering}
\caption{Grain boundary segregation $\Gamma$ (a), dynamic grain boundary free
energy $\gamma$ (b), and entropy production rate $\sigma$ (c) as
functions of the grain boundary speed $V$ for the alloy composition
$c=0.1$ and temperature $T=0.8$ (alloy A in Fig.~\ref{fig:Phase_diag}(b)).
The red and blue curves were obtained by gradually increasing and
decreasing the speed, respectively.\label{fig:Properties_A}}

\end{figure}

\begin{figure}
\noindent \begin{centering}
\textbf{(a)}\enskip{}\includegraphics[width=0.42\textwidth]{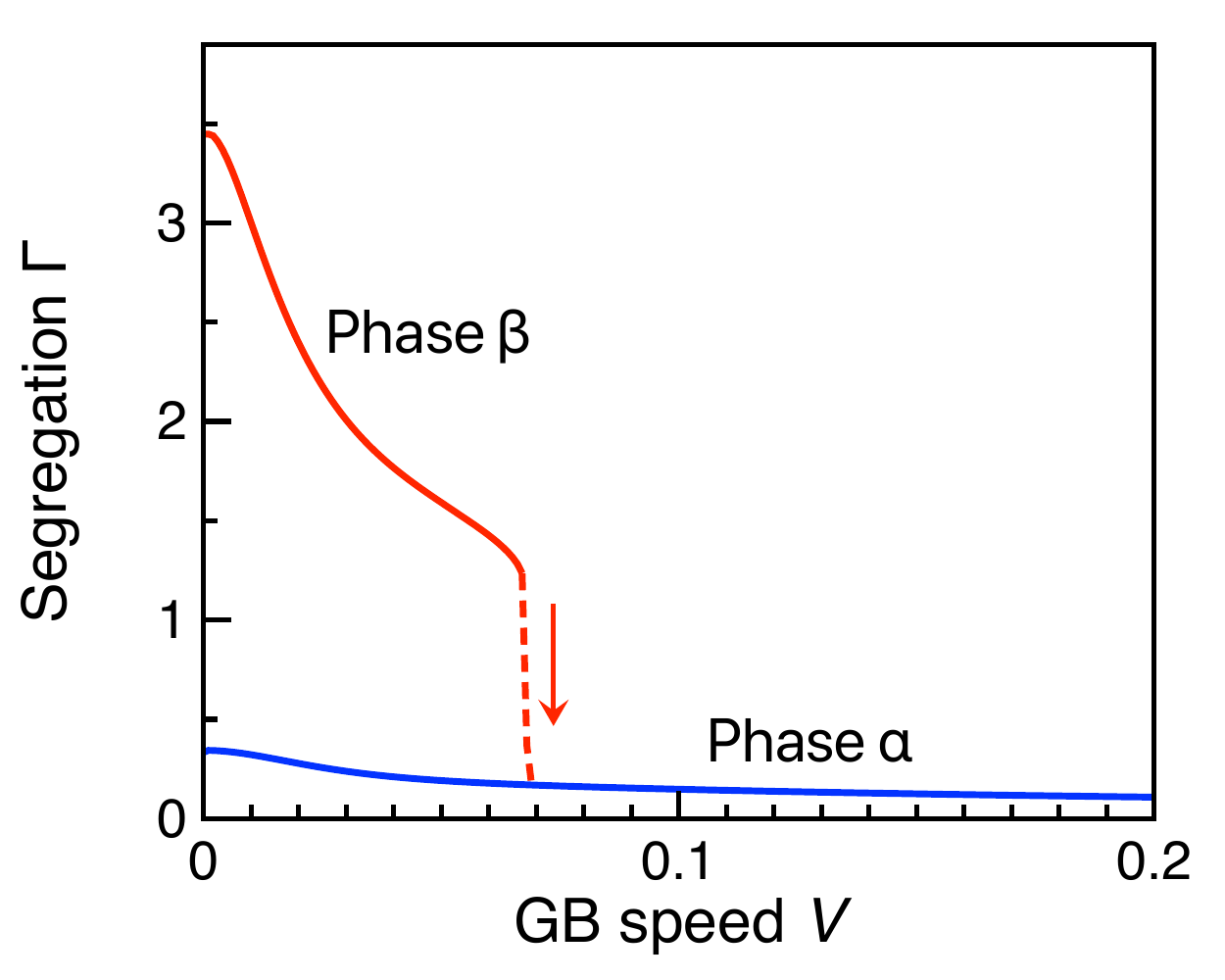}
\par\end{centering}
\noindent \begin{centering}
\bigskip{}
\par\end{centering}
\noindent \begin{centering}
\textbf{(b)}\enskip{}\includegraphics[width=0.42\textwidth]{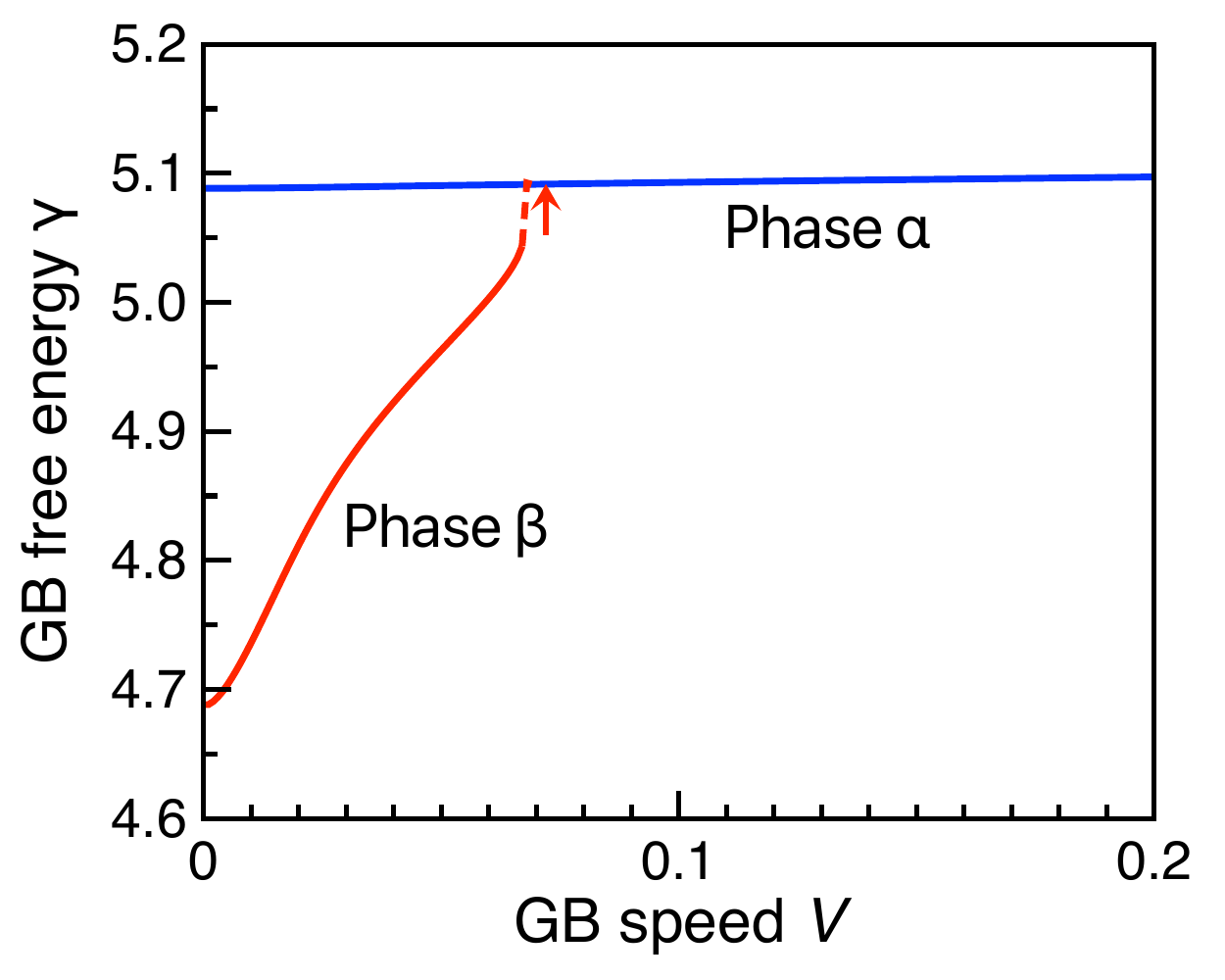}
\par\end{centering}
\noindent \begin{centering}
\bigskip{}
\par\end{centering}
\noindent \begin{centering}
\textbf{(c)}\enskip{}\includegraphics[width=0.42\textwidth]{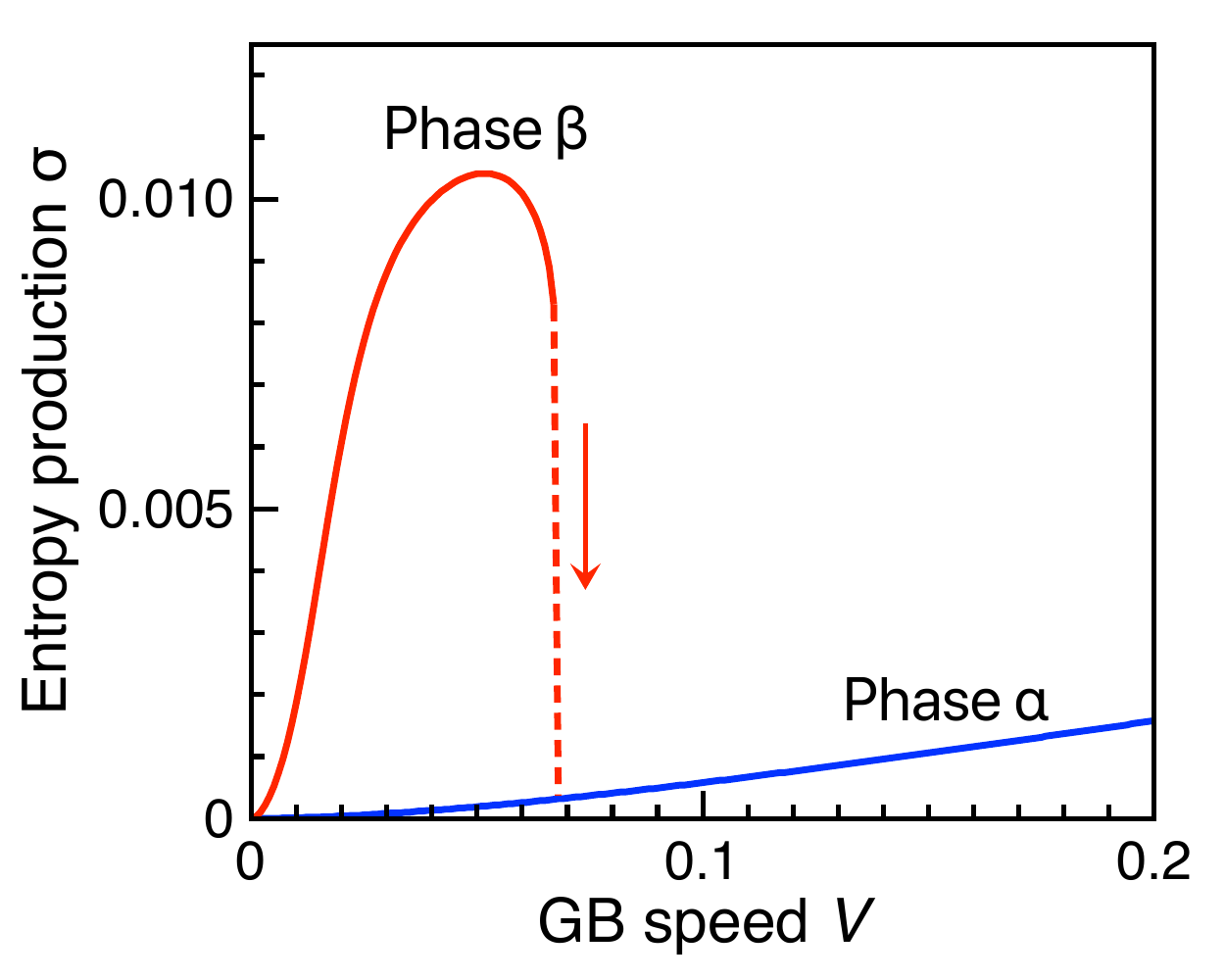}
\par\end{centering}
\caption{Grain boundary segregation $\Gamma$ (a), dynamic grain boundary free
energy $\gamma$ (b), and entropy production rate $\sigma$ (c) as
functions of the grain boundary speed $V$ for the alloy composition
$c=0.08$ and temperature $T=0.8$. The red and blue curves were obtained
by, respectively, increasing and decreasing the speed.\label{fig:Properties_B-1}}
\end{figure}

The second unresolved issue is the lack of a criterion for predicting
the transformations between the dynamic grain boundary phases. For
equilibrium grain boundaries, the most stable phase is the one which
minimizes $\gamma$ \citep{Willard_Gibbs,Frolov:2015ab}. Coexisting
phases have equal values of $\gamma$. One would hope that the dynamic
extension of $\gamma$ defined by Eq.(\ref{eq:gamma}) could play
the same role as the equilibrium $\gamma$ in dictating the phase
selection rules for moving grain boundaries. Calculations show that
this assumption is not valid. This is clear from the observation that
the $\gamma$-plots of the dynamic phases do not cross within the
hysteresis region, in which the phases are expected to reach two-phase
coexistence (see Figs.~\ref{fig:Properties_A} and \ref{fig:Properties_B-1}),
as they do in the equilibrium case (cf.~Fig.~\ref{fig:gamma_plot}).
This is the reason why we were unable to construct the dynamic continuation
of the phase coexistence line on the grain boundary phase diagram.
We hypothesize that a different grain boundary property might exist
that should be minimized to predict the dynamic phase transformations,
and whose values are equal in coexisting phases. The well-known extremum
principles of non-equilibrium thermodynamics \citep{Onsager1931a,Prigogine1968,Ziegler:1987aa,Hillert:2006aa,Martyushev:2006aa,Svoboda:2011aa,Fischer:2014aa}
do not provide an answer. Such principles only predict that a steady
state corresponds to a local extremum (maximum or minimum) of an appropriate
function, such as the entropy production rate (respectively, the negative
of the free energy rate under isothermal conditions). However, when
several different steady states can be reached by the system, a global
extremum principle is required to make a selection. At present, we
are not aware of a criterion that would permit phase stability predictions
for moving grain boundaries.

\bigskip{}

\noindent \textbf{Acknowledgements}

This work was supported by the U.S.~Army Research Office under a
contract number W911NF-15-1-0077.


\begin{thebibliography}{58}
\expandafter\ifx\csname natexlab\endcsname\relax\def\natexlab#1{#1}\fi
\providecommand{\url}[1]{\texttt{#1}}
\providecommand{\href}[2]{#2}
\providecommand{\path}[1]{#1}
\providecommand{\DOIprefix}{doi:}
\providecommand{\ArXivprefix}{arXiv:}
\providecommand{\URLprefix}{URL: }
\providecommand{\Pubmedprefix}{pmid:}
\providecommand{\doi}[1]{\href{http://dx.doi.org/#1}{\path{#1}}}
\providecommand{\Pubmed}[1]{\href{pmid:#1}{\path{#1}}}
\providecommand{\bibinfo}[2]{#2}
\ifx\xfnm\relax \def\xfnm[#1]{\unskip,\space#1}\fi
\bibitem[{Sutton and Balluffi(1995)}]{Balluffi95}
\bibinfo{author}{A.~P. Sutton}, \bibinfo{author}{R.~W. Balluffi},
  \bibinfo{title}{Interfaces in Crystalline Materials},
  \bibinfo{publisher}{Clarendon Press}, \bibinfo{address}{Oxford},
  \bibinfo{year}{1995}.
\bibitem[{Cahn(1962)}]{Cahn1962}
\bibinfo{author}{J.~W. Cahn},
\newblock \bibinfo{title}{The impurity-drag effect in grain boundary motion},
\newblock \bibinfo{journal}{Acta Mater.} \bibinfo{volume}{10}
  (\bibinfo{year}{1962}) \bibinfo{pages}{789--798}.
\bibitem[{L{\"u}cke and St{\"u}we(1963)}]{Lucke-Stuwe-1963}
\bibinfo{author}{K.~L{\"u}cke}, \bibinfo{author}{H.~P. St{\"u}we},
\newblock in: \bibinfo{editor}{L.~Himmel} (Ed.), \bibinfo{booktitle}{Recovery
  and Recrystallization of Metals}, \bibinfo{publisher}{Interscience
  Publishers}, \bibinfo{address}{New York}, \bibinfo{year}{1963}, pp.
  \bibinfo{pages}{171--210}.
\bibitem[{L{\"u}cke and St{\"u}we(1971)}]{Lucke:1971aa}
\bibinfo{author}{K.~L{\"u}cke}, \bibinfo{author}{H.~P. St{\"u}we},
\newblock \bibinfo{title}{On the theory of impurity controlled grain boundary
  motion},
\newblock \bibinfo{journal}{Acta Metall.} \bibinfo{volume}{19}
  (\bibinfo{year}{1971}) \bibinfo{pages}{1087--1099}.
\bibitem[{Ma et~al.(2003)Ma, Dregia, and Wang}]{Wang03}
\bibinfo{author}{N.~Ma}, \bibinfo{author}{S.~A. Dregia},
  \bibinfo{author}{Y.~Wang},
\newblock \bibinfo{title}{Solute segregation transition and drag force on grain
  boundaries},
\newblock \bibinfo{journal}{Acta Mater.} \bibinfo{volume}{51}
  (\bibinfo{year}{2003}) \bibinfo{pages}{3687--3700}.
\bibitem[{Li et~al.(2009)Li, Wang, and Yang}]{Li:2009aa}
\bibinfo{author}{J.~Li}, \bibinfo{author}{J.~Wang}, \bibinfo{author}{G.~Yang},
\newblock \bibinfo{title}{Phase field modeling of grain boundary migration with
  solute drag},
\newblock \bibinfo{journal}{Acta Mater.} \bibinfo{volume}{57}
  (\bibinfo{year}{2009}) \bibinfo{pages}{2108--2120}.
\bibitem[{Shahandeh et~al.(2012)Shahandeh, Greenwood, and
  Militzer}]{Shahandeh:2012aa}
\bibinfo{author}{S.~Shahandeh}, \bibinfo{author}{M.~Greenwood},
  \bibinfo{author}{M.~Militzer},
\newblock \bibinfo{title}{Friction pressure method for simulating solute drag
  and particle pinning in a multiphase-field model},
\newblock \bibinfo{journal}{Model. Simul. Mater. Sci. Eng.}
  \bibinfo{volume}{20} (\bibinfo{year}{2012}) \bibinfo{pages}{065008}.
\bibitem[{Gr{\"o}nhagen and Agren(2007)}]{Gronhagen:2007aa}
\bibinfo{author}{K.~Gr{\"o}nhagen}, \bibinfo{author}{J.~Agren},
\newblock \bibinfo{title}{Grain-boundary segregation and dynamic solute drag
  theory --- {A} phase-field approach},
\newblock \bibinfo{journal}{Acta Mater.} \bibinfo{volume}{55}
  (\bibinfo{year}{2007}) \bibinfo{pages}{955--960}.
\bibitem[{Abdeljawad et~al.(2017)Abdeljawad, Lu, Argibay, Clark, Boyce, and
  Foiles}]{Abdeljawad:2017aa}
\bibinfo{author}{F.~Abdeljawad}, \bibinfo{author}{P.~Lu},
  \bibinfo{author}{N.~Argibay}, \bibinfo{author}{B.~G. Clark},
  \bibinfo{author}{B.~L. Boyce}, \bibinfo{author}{S.~M. Foiles},
\newblock \bibinfo{title}{Grain boundary segregation in immiscible
  nanocrystalline alloys},
\newblock \bibinfo{journal}{Acta Mater.} \bibinfo{volume}{126}
  (\bibinfo{year}{2017}) \bibinfo{pages}{528--539}.
\bibitem[{Greenwood et~al.(2012)Greenwood, Sinclair, and
  Militzer}]{Greenwood:2012aa}
\bibinfo{author}{M.~Greenwood}, \bibinfo{author}{C.~Sinclair},
  \bibinfo{author}{M.~Militzer},
\newblock \bibinfo{title}{Phase field crystal model of solute drag},
\newblock \bibinfo{journal}{Acta Mater.} \bibinfo{volume}{60}
  (\bibinfo{year}{2012}) \bibinfo{pages}{5752--5761}.
\bibitem[{Mendelev and Srolovitz(2002)}]{Mendelev02a}
\bibinfo{author}{M.~I. Mendelev}, \bibinfo{author}{D.~J. Srolovitz},
\newblock \bibinfo{title}{Impurity effects on grain boundary migration},
\newblock \bibinfo{journal}{Model. Simul. Mater. Sci. Eng.}
  \bibinfo{volume}{10} (\bibinfo{year}{2002}) \bibinfo{pages}{R79--R109}.
\bibitem[{Sun and Deng(2014)}]{Sun:2014aa}
\bibinfo{author}{H.~Sun}, \bibinfo{author}{C.~Deng},
\newblock \bibinfo{title}{Direct quantification of solute effects on grain
  boundary motion by atomistic simulations},
\newblock \bibinfo{journal}{Comp. Mater. Sci.} \bibinfo{volume}{93}
  (\bibinfo{year}{2014}) \bibinfo{pages}{137--143}.
\bibitem[{Wicaksono et~al.(2013)Wicaksono, Sinclair, and
  Militzer}]{Wicaksono:2013aa}
\bibinfo{author}{A.~T. Wicaksono}, \bibinfo{author}{C.~W. Sinclair},
  \bibinfo{author}{M.~Militzer},
\newblock \bibinfo{title}{A three-dimensional atomistic kinetic {Monte Carlo}
  study of dynamic solute-interface interaction},
\newblock \bibinfo{journal}{Model. Simul. Mater. Sci. Eng.}
  \bibinfo{volume}{21} (\bibinfo{year}{2013}) \bibinfo{pages}{085010}.
\bibitem[{Mendelev et~al.(2001)Mendelev, Srolovitz, and E}]{Mendelev:2001aa}
\bibinfo{author}{M.~I. Mendelev}, \bibinfo{author}{D.~J. Srolovitz},
  \bibinfo{author}{W.~E},
\newblock \bibinfo{title}{Grain-boundary migration in the presence of diffusing
  impurities: simulations and analytical models},
\newblock \bibinfo{journal}{Philos. Mag.} \bibinfo{volume}{81}
  (\bibinfo{year}{2001}) \bibinfo{pages}{2243--2269}.
\bibitem[{Rahman et~al.(2016)Rahman, Zurob, and Hoyt}]{Rahman:2016aa}
\bibinfo{author}{M.~J. Rahman}, \bibinfo{author}{H.~S. Zurob},
  \bibinfo{author}{J.~J. Hoyt},
\newblock \bibinfo{title}{Molecular dynamics study of solute pinning effects on
  grain boundary migration in the aluminum magnesium alloy system},
\newblock \bibinfo{journal}{Metall. Mater. Trans. A} \bibinfo{volume}{47}
  (\bibinfo{year}{2016}) \bibinfo{pages}{1889--1897}.
\bibitem[{Kim and Park(2008)}]{Kim:2008aa}
\bibinfo{author}{S.~G. Kim}, \bibinfo{author}{Y.~B. Park},
\newblock \bibinfo{title}{Grain boundary segregation, solute drag and abnormal
  grain growth},
\newblock \bibinfo{journal}{Acta Mater.} \bibinfo{volume}{56}
  (\bibinfo{year}{2008}) \bibinfo{pages}{3739--3753}.
\bibitem[{Mishin et~al.(2010)Mishin, Asta, and Li}]{Mishin2010a}
\bibinfo{author}{Y.~Mishin}, \bibinfo{author}{M.~Asta},
  \bibinfo{author}{J.~Li},
\newblock \bibinfo{title}{Atomistic modeling of interfaces and their impact on
  microstructure and properties},
\newblock \bibinfo{journal}{Acta Mater.} \bibinfo{volume}{58}
  (\bibinfo{year}{2010}) \bibinfo{pages}{1117 -- 1151}.
\bibitem[{Mishin(2015)}]{Mishin:2015ac}
\bibinfo{author}{Y.~Mishin},
\newblock \bibinfo{title}{An atomistic view of grain boundary diffusion},
\newblock \bibinfo{journal}{Defect and Diffusion Forum} \bibinfo{volume}{363}
  (\bibinfo{year}{2015}) \bibinfo{pages}{1--11}.
\bibitem[{Cantwell et~al.(2013)Cantwell, Tang, Dillon, Luo, Rohrer, and
  Harmer}]{Cantwell-2013}
\bibinfo{author}{P.~R. Cantwell}, \bibinfo{author}{M.~Tang},
  \bibinfo{author}{S.~J. Dillon}, \bibinfo{author}{J.~Luo},
  \bibinfo{author}{G.~S. Rohrer}, \bibinfo{author}{M.~P. Harmer},
\newblock \bibinfo{title}{Grain boundary complexions},
\newblock \bibinfo{journal}{Acta Mater.} \bibinfo{volume}{62}
  (\bibinfo{year}{2013}) \bibinfo{pages}{1--48}.
\bibitem[{Baram et~al.(2011)Baram, Chatain, and Kaplan}]{Baram08042011}
\bibinfo{author}{M.~Baram}, \bibinfo{author}{D.~Chatain},
  \bibinfo{author}{W.~D. Kaplan},
\newblock \bibinfo{title}{Nanometer-thick equilibrium films: The interface
  between thermodynamics and atomistics},
\newblock \bibinfo{journal}{Science} \bibinfo{volume}{332}
  (\bibinfo{year}{2011}) \bibinfo{pages}{206--209}.
\bibitem[{Luo et~al.(2011)Luo, Cheng, Asl, Kiely, and Harmer}]{Luo23092011}
\bibinfo{author}{J.~Luo}, \bibinfo{author}{H.~Cheng}, \bibinfo{author}{K.~M.
  Asl}, \bibinfo{author}{C.~J. Kiely}, \bibinfo{author}{M.~P. Harmer},
\newblock \bibinfo{title}{The role of a bilayer interfacial phase on liquid
  metal embrittlement},
\newblock \bibinfo{journal}{Science} \bibinfo{volume}{333}
  (\bibinfo{year}{2011}) \bibinfo{pages}{1730--1733}.
\bibitem[{Harmer(2011)}]{Harmer08042011}
\bibinfo{author}{M.~P. Harmer},
\newblock \bibinfo{title}{The phase behavior of interfaces},
\newblock \bibinfo{journal}{Science} \bibinfo{volume}{332}
  (\bibinfo{year}{2011}) \bibinfo{pages}{182--183}.
\bibitem[{Rheinheimer and Hoffmann(2015)}]{Rheinheimer201568}
\bibinfo{author}{W.~Rheinheimer}, \bibinfo{author}{M.~J. Hoffmann},
\newblock \bibinfo{title}{Non-arrhenius behavior of grain growth in strontium
  titanate: New evidence for a structural transition of grain boundaries},
\newblock \bibinfo{journal}{Scripta Mater.} \bibinfo{volume}{101}
  (\bibinfo{year}{2015}) \bibinfo{pages}{68 -- 71}.
\bibitem[{Hart(1972)}]{Hart:1972aa}
\bibinfo{author}{E.~W. Hart},
\newblock \bibinfo{title}{Grain boundary phase transformations},
\newblock in: \bibinfo{editor}{H.~Hu} (Ed.), \bibinfo{booktitle}{Nature and
  behavior of grain boundaries}, \bibinfo{publisher}{Plenum},
  \bibinfo{address}{New York}, \bibinfo{year}{1972}, pp.
  \bibinfo{pages}{155--170}.
\bibitem[{Cahn(1982)}]{Cahn82a}
\bibinfo{author}{J.~W. Cahn},
\newblock \bibinfo{title}{Transitions and phase equilibria among grain boundary
  structures},
\newblock \bibinfo{journal}{J. Physique Colloques} \bibinfo{volume}{43}
  (\bibinfo{year}{1982}) \bibinfo{pages}{199--213}.
\bibitem[{Rottman(1988)}]{Rottman1988a}
\bibinfo{author}{C.~Rottman},
\newblock \bibinfo{title}{Theory of phase transitions at internal interfaces},
\newblock \bibinfo{journal}{J. de Physique Colloque} \bibinfo{volume}{49}
  (\bibinfo{year}{1988}) \bibinfo{pages}{313--322}.
\bibitem[{Frolov and Mishin(2015)}]{Frolov:2015ab}
\bibinfo{author}{T.~Frolov}, \bibinfo{author}{Y.~Mishin},
\newblock \bibinfo{title}{Phases, phase equilibria, and phase rules in
  low-dimensional systems},
\newblock \bibinfo{journal}{J. Chem. Phys.} \bibinfo{volume}{143}
  (\bibinfo{year}{2015}) \bibinfo{pages}{044706}.
\bibitem[{Frolov et~al.(2013{\natexlab{a}})Frolov, Olmsted, Asta, and
  Mishin}]{Frolov2013}
\bibinfo{author}{T.~Frolov}, \bibinfo{author}{D.~L. Olmsted},
  \bibinfo{author}{M.~Asta}, \bibinfo{author}{Y.~Mishin},
\newblock \bibinfo{title}{Structural phase transformations in metallic grain
  boundaries},
\newblock \bibinfo{journal}{Nature Communications} \bibinfo{volume}{4}
  (\bibinfo{year}{2013}{\natexlab{a}}) \bibinfo{pages}{1899}.
\bibitem[{Frolov et~al.(2013{\natexlab{b}})Frolov, Divinski, Asta, and
  Mishin}]{Frolov2013a}
\bibinfo{author}{T.~Frolov}, \bibinfo{author}{S.~V. Divinski},
  \bibinfo{author}{M.~Asta}, \bibinfo{author}{Y.~Mishin},
\newblock \bibinfo{title}{Effect of interface phase transformations on
  diffusion and segregation in high-angle grain boundaries},
\newblock \bibinfo{journal}{Phys. Rev. Lett.} \bibinfo{volume}{110}
  (\bibinfo{year}{2013}{\natexlab{b}}) \bibinfo{pages}{255502}.
\bibitem[{Frolov et~al.(2015)Frolov, M.Asta, and Mishin}]{Frolov:2015aa}
\bibinfo{author}{T.~Frolov}, \bibinfo{author}{M.Asta},
  \bibinfo{author}{Y.~Mishin},
\newblock \bibinfo{title}{Segregation-induced phase transformations in grain
  boundaries},
\newblock \bibinfo{journal}{Phys. Rev. {\rm B}} \bibinfo{volume}{92}
  (\bibinfo{year}{2015}) \bibinfo{pages}{020103(R)}.
\bibitem[{Frolov et~al.(2016)Frolov, Asta, and Mishin}]{Frolov:2016aa}
\bibinfo{author}{T.~Frolov}, \bibinfo{author}{M.~Asta},
  \bibinfo{author}{Y.~Mishin},
\newblock \bibinfo{title}{Phase transformations at interfaces: Observations
  from atomistic modeling},
\newblock \bibinfo{journal}{Current Opinion in Solid State and Materials
  Science} \bibinfo{volume}{20} (\bibinfo{year}{2016})
  \bibinfo{pages}{308--315.}
\bibitem[{Hickman and Mishin(2017)}]{Hickman__2017a}
\bibinfo{author}{J.~Hickman}, \bibinfo{author}{Y.~Mishin},
\newblock \bibinfo{title}{Extra variable in grain boundary description},
\newblock \bibinfo{journal}{Phys. Rev. Materials} \bibinfo{volume}{1}
  (\bibinfo{year}{2017}) \bibinfo{pages}{010601}.
\bibitem[{Yang et~al.(2018)Yang, Zhou, Zheng, Ong, and Luo}]{Yang:2018ab}
\bibinfo{author}{S.~Yang}, \bibinfo{author}{N.~Zhou},
  \bibinfo{author}{H.~Zheng}, \bibinfo{author}{S.~Ong},
  \bibinfo{author}{J.~Luo},
\newblock \bibinfo{title}{First-order interfacial transformations with a
  critical point: {Breaking} the symmetry at a symmetric tilt grain boundary},
\newblock \bibinfo{journal}{Phys. Rev. Lett.} \bibinfo{volume}{120}
  (\bibinfo{year}{2018}) \bibinfo{pages}{085702}.
\bibitem[{Frolov et~al.(2018)Frolov, Zhu, Oppelstrup, Marian, and
  Rudd}]{Frolov:2018aa}
\bibinfo{author}{T.~Frolov}, \bibinfo{author}{Q.~Zhu},
  \bibinfo{author}{T.~Oppelstrup}, \bibinfo{author}{J.~Marian},
  \bibinfo{author}{R.~E. Rudd},
\newblock \bibinfo{title}{Structures and transitions in bcc tungsten grain
  boundaries and their role in the absorption of point defects},
\newblock \bibinfo{journal}{Acta Mater.} \bibinfo{volume}{159}
  (\bibinfo{year}{2018}) \bibinfo{pages}{123--134}.
\bibitem[{Wynblatt and Chatain(2006)}]{Wynblatt2006}
\bibinfo{author}{P.~Wynblatt}, \bibinfo{author}{D.~Chatain},
\newblock \bibinfo{title}{Anisotropy of segregation at grain boundaries and
  surfaces},
\newblock \bibinfo{journal}{Metall. Mater. Trans. {\rm A}} \bibinfo{volume}{37}
  (\bibinfo{year}{2006}) \bibinfo{pages}{2595--2620}.
\bibitem[{Wynblatt et~al.(2006)Wynblatt, Chatain, and Peng}]{Wynblatt:2006aa}
\bibinfo{author}{P.~Wynblatt}, \bibinfo{author}{D.~Chatain},
  \bibinfo{author}{Y.~Peng},
\newblock \bibinfo{title}{Some aspects of the anisotropy of grain boundary
  segregation and wetting},
\newblock \bibinfo{journal}{J. Mate. Sci.} \bibinfo{volume}{41}
  (\bibinfo{year}{2006}) \bibinfo{pages}{7760--7768}.
\bibitem[{Wynblatt(2008)}]{Wynblatt2008}
\bibinfo{author}{P.~Wynblatt},
\newblock \bibinfo{title}{Interfacial segregation effects in wetting
  phenomena},
\newblock \bibinfo{journal}{Annu. Rev. Mater. Res.} \bibinfo{volume}{38}
  (\bibinfo{year}{2008}) \bibinfo{pages}{173--196}.
\bibitem[{Wynblatt and Chatain(2008)}]{Wynblatt:2008aa}
\bibinfo{author}{P.~Wynblatt}, \bibinfo{author}{D.~Chatain},
\newblock \bibinfo{title}{Solid-state wetting transitions at grain boundaries},
\newblock \bibinfo{journal}{Mater. Sci. Eng. {\rm A}} \bibinfo{volume}{495}
  (\bibinfo{year}{2008}) \bibinfo{pages}{119--125}.
\bibitem[{Larch{\'e} and Cahn(1978)}]{Larche_Cahn_78}
\bibinfo{author}{F.~C. Larch{\'e}}, \bibinfo{author}{J.~W. Cahn},
\newblock \bibinfo{title}{Thermodynamic equilibrium of multiphase solids under
  stress},
\newblock \bibinfo{journal}{Acta Metall.} \bibinfo{volume}{26}
  (\bibinfo{year}{1978}) \bibinfo{pages}{1579--1589}.
\bibitem[{Larch{\'e} and Cahn(1985)}]{Larche1985}
\bibinfo{author}{F.~C. Larch{\'e}}, \bibinfo{author}{J.~W. Cahn},
\newblock \bibinfo{title}{The interactions of composition and stress in
  crystalline solids},
\newblock \bibinfo{journal}{Acta Metall.} \bibinfo{volume}{33}
  (\bibinfo{year}{1985}) \bibinfo{pages}{331--367}.
\bibitem[{Hillert and Sundman(1976)}]{Hillert:1976aa}
\bibinfo{author}{M.~Hillert}, \bibinfo{author}{B.~Sundman},
\newblock \bibinfo{title}{A treatment of the solute drag on moving grain
  boundaries and phase interfaces in binary alloys},
\newblock \bibinfo{journal}{Acta Metall.} \bibinfo{volume}{24}
  (\bibinfo{year}{1976}) \bibinfo{pages}{731--743}.
\bibitem[{Hillert(1999)}]{Hillert:1999aa}
\bibinfo{author}{M.~Hillert},
\newblock \bibinfo{title}{Solute drag, solute trapping and diffusional
  dissipation of {Gibbs} energy},
\newblock \bibinfo{journal}{Acta Mater.} \bibinfo{volume}{47}
  (\bibinfo{year}{1999}) \bibinfo{pages}{4481--4505}.
\bibitem[{Hillert et~al.(2001)Hillert, Odquist, and Agren}]{Hillert:2001aa}
\bibinfo{author}{M.~Hillert}, \bibinfo{author}{J.~Odquist},
  \bibinfo{author}{J.~Agren},
\newblock \bibinfo{title}{Comparison between solute drag and dissipation of
  gibbs energy by diffusion},
\newblock \bibinfo{journal}{Scripta Mater.} \bibinfo{volume}{45}
  (\bibinfo{year}{2001}) \bibinfo{pages}{221--227}.
\bibitem[{Prigogine(1968)}]{Prigogine1968}
\bibinfo{author}{I.~Prigogine}, \bibinfo{title}{Introduction to thermodynamics
  of irreversible processes}, \bibinfo{publisher}{Interscience Publishers},
  \bibinfo{address}{New York}, \bibinfo{year}{1968}.
\bibitem[{\mbox{De Groot} and Mazur(1984)}]{De-Groot1984}
\bibinfo{author}{S.~R. \mbox{De Groot}}, \bibinfo{author}{P.~Mazur},
  \bibinfo{title}{Non-equilibrium thermodynamics}, \bibinfo{publisher}{Dover},
  \bibinfo{address}{New York}, \bibinfo{year}{1984}.
\bibitem[{Gibbs(1948)}]{Willard_Gibbs}
\bibinfo{author}{J.~W. Gibbs},
\newblock \bibinfo{title}{On the equilibrium of heterogeneous substances},
\newblock in: \bibinfo{booktitle}{The collected works of J. W. Gibbs},
  volume~\bibinfo{volume}{1}, \bibinfo{publisher}{Yale University Press},
  \bibinfo{address}{New Haven}, \bibinfo{year}{1948}, pp.
  \bibinfo{pages}{55--349}.
\bibitem[{Carroll(1865)}]{Carroll:1865aa}
\bibinfo{author}{L.~Carroll}, \bibinfo{title}{{Alice's Adventures in
  Wonderland}}, \bibinfo{publisher}{Mcmillan}, \bibinfo{address}{London, UK},
  \bibinfo{year}{1865}.
\bibitem[{Roy and Bauer(1975)}]{Roy:1975aa}
\bibinfo{author}{A.~Roy}, \bibinfo{author}{C.~L. Bauer},
\newblock \bibinfo{title}{Effect of impurities on the stability of moving grain
  boundary},
\newblock \bibinfo{journal}{Acta Mater.} \bibinfo{volume}{23}
  (\bibinfo{year}{1975}) \bibinfo{pages}{957--963}.
\bibitem[{Korzhenevskii et~al.(2002)Korzhenevskii, R, and
  Schmitz}]{Korzhenevskii:2002aa}
\bibinfo{author}{A.~L. Korzhenevskii}, \bibinfo{author}{B.~R},
  \bibinfo{author}{R.~Schmitz},
\newblock \bibinfo{title}{Morphological instabilities of moving extended
  defects},
\newblock \bibinfo{journal}{Europhys. Lett.} \bibinfo{volume}{59}
  (\bibinfo{year}{2002}) \bibinfo{pages}{533--539}.
\bibitem[{Korzhenevskii et~al.(2006)Korzhenevskii, R, and
  Schmitz}]{Korzhenevskii:2006aa}
\bibinfo{author}{A.~L. Korzhenevskii}, \bibinfo{author}{B.~R},
  \bibinfo{author}{R.~Schmitz},
\newblock \bibinfo{title}{Stability of grain-boundary motion in the presence of
  impurities},
\newblock \bibinfo{journal}{Acta Mater.} \bibinfo{volume}{54}
  (\bibinfo{year}{2006}) \bibinfo{pages}{1595--1596}.
\bibitem[{Mullins and Sekerka(1964)}]{Mullins:1964aa}
\bibinfo{author}{W.~W. Mullins}, \bibinfo{author}{R.~F. Sekerka},
\newblock \bibinfo{title}{Stability of a planar interface during solidification
  of a dilute binary alloy},
\newblock \bibinfo{journal}{J. Appl. Phys.} \bibinfo{volume}{35}
  (\bibinfo{year}{1964}) \bibinfo{pages}{444--451}.
\bibitem[{Herring(1951)}]{Herring:1951aa}
\bibinfo{author}{C.~Herring},
\newblock \bibinfo{title}{Surface tension as a motivation for sintering},
\newblock in: \bibinfo{booktitle}{The Physics of Powder Metallurgy},
  \bibinfo{publisher}{McGraw-Hill}, \bibinfo{address}{New York},
  \bibinfo{year}{1951}, pp. \bibinfo{pages}{143--179}.
\bibitem[{Onsager(1931)}]{Onsager1931a}
\bibinfo{author}{L.~Onsager},
\newblock \bibinfo{title}{Reciprocal relations in irreversible processes. {I}},
\newblock \bibinfo{journal}{Phys. Rev.} \bibinfo{volume}{37}
  (\bibinfo{year}{1931}) \bibinfo{pages}{405--426}.
\bibitem[{Ziegler and Wehri(1987)}]{Ziegler:1987aa}
\bibinfo{author}{H.~Ziegler}, \bibinfo{author}{C.~Wehri},
\newblock \bibinfo{title}{On a principle of maximal rate of entropy
  production},
\newblock \bibinfo{journal}{J. Non-Equilib. Thermodyn.} \bibinfo{volume}{12}
  (\bibinfo{year}{1987}) \bibinfo{pages}{229--244}.
\bibitem[{Hillert and Agren(2006)}]{Hillert:2006aa}
\bibinfo{author}{M.~Hillert}, \bibinfo{author}{J.~Agren},
\newblock \bibinfo{title}{Extremum principles for irreversible processes},
\newblock \bibinfo{journal}{Acta Mater.} \bibinfo{volume}{54}
  (\bibinfo{year}{2006}) \bibinfo{pages}{2063--2066}.
\bibitem[{Martyushev and Seleznev(2006)}]{Martyushev:2006aa}
\bibinfo{author}{L.~M. Martyushev}, \bibinfo{author}{V.~D. Seleznev},
\newblock \bibinfo{title}{Maximum entropy production principle in physics,
  chemistry and biology},
\newblock \bibinfo{journal}{Physics Reports} \bibinfo{volume}{426}
  (\bibinfo{year}{2006}) \bibinfo{pages}{1--45}.
\bibitem[{Svoboda et~al.(2011)Svoboda, Fischer, and Leindl}]{Svoboda:2011aa}
\bibinfo{author}{J.~Svoboda}, \bibinfo{author}{F.~D. Fischer},
  \bibinfo{author}{M.~Leindl},
\newblock \bibinfo{title}{Transient solute drag in migrating grain boundaries},
\newblock \bibinfo{journal}{Acta Mater.} \bibinfo{volume}{59}
  (\bibinfo{year}{2011}) \bibinfo{pages}{6556--6562}.
\bibitem[{Fischer et~al.(2014)Fischer, Svoboda, and Petryk}]{Fischer:2014aa}
\bibinfo{author}{F.~D. Fischer}, \bibinfo{author}{J.~Svoboda},
  \bibinfo{author}{H.~Petryk},
\newblock \bibinfo{title}{Thermodynamic extremal principles for irreversible
  processes in materials science},
\newblock \bibinfo{journal}{Acta Mater.} \bibinfo{volume}{67}
  (\bibinfo{year}{2014}) \bibinfo{pages}{1--20}.

\end{thebibliography}

\newpage\clearpage{}

\begin{table}
\begin{tabular}{ccccccccccccc}
\hline 
Physical &  & Dimensionless &  &  & Physical &  & Dimensionless &  &  & Physical &  & Dimensionless\tabularnewline
\hline 
 &  &  &  &  &  &  &  &  &  &  &  & \tabularnewline
$x$ &  & $\dfrac{x}{a}$ &  &  & $\omega_{i}$, $\omega_{i}^{\prime}$ &  & $\dfrac{\omega_{i}}{kT_{c}}$, $\dfrac{\omega_{i}^{\prime}}{kT_{c}}$ &  &  & $\Gamma$ &  & $s\Gamma$\tabularnewline
 &  &  &  &  &  &  &  &  &  &  &  & \tabularnewline
$\lambda$ &  & $\dfrac{\lambda}{a}$ &  &  & $\mu_{i}$ &  & $\dfrac{\mu_{i}}{kT_{c}}$ &  &  & $\gamma$ &  & $\dfrac{s\gamma}{kT_{c}}$\tabularnewline
 &  &  &  &  &  &  &  &  &  &  &  & \tabularnewline
$t$ &  & $\dfrac{Dt}{a^{2}}$ &  &  & $\mu$ &  & $\dfrac{\mu}{kT_{c}}$ &  &  & $f$ &  & $\dfrac{saf}{kT_{c}}$\tabularnewline
 &  &  &  &  &  &  &  &  &  &  &  & \tabularnewline
$V$ &  & $\dfrac{aV}{D}$ &  &  & $\varphi$ &  & $\dfrac{\varphi}{kT_{c}}$ &  &  & $J_{i}$ &  & $\dfrac{sa^{2}J_{i}}{D}$\tabularnewline
 &  &  &  &  &  &  &  &  &  &  &  & \tabularnewline
$T$ &  & $\dfrac{T}{T_{c}}$ &  &  & $F$ &  & $\dfrac{sF}{kT_{c}}$ &  &  & $M_{i}$ &  & $\dfrac{sa^{2}kT_{c}M_{i}}{D}$\tabularnewline
 &  &  &  &  &  &  &  &  &  &  &  & \tabularnewline
$\varepsilon_{i}^{\nu}$, $\varepsilon_{i}^{\prime\nu}$ &  & $\dfrac{\varepsilon_{i}^{\nu}}{kT_{c}}$, $\dfrac{\varepsilon_{i}^{\prime\nu}}{kT_{c}}$ &  &  & $\sigma$ &  & $\dfrac{s\sigma a^{2}}{kD}$ &  &  & $M$ &  & $\dfrac{sa^{2}kT_{c}M}{D}$\tabularnewline
\hline 
\end{tabular}

\caption{Dimensionless parameters of the model used for numerical calculations.\label{tab:Dimensionless-parameters}}
\end{table}

\newpage\clearpage{}

\end{document}